# Revisiting X-ray-Bright-Optically-Normal-Galaxies with the Chandra Source Catalog

Dong-Woo Kim[1], Amanda Malnati[2], Alyssa Cassity[3],
Giuseppina Fabbiano[1], Juan Rafael Martinez Galarza[1], Ewan O'Sullivan[1]

1. Center for Astrophysics | Harvard and Smithsonian
2. Tufts University
3. University of British Columbia

(July 28, 2023)

## Abstract

X-ray bright optically normal galaxies (XBONGs) are galaxies with X-ray luminosities consistent with those of active galactic nuclei (AGNs) but no evidence of AGN optical emission lines. Crossmatching the Chandra Source Catalog version 2 (CSC2) with the Sloan Digital Sky Survey (SDSS) sample of spectroscopically classified galaxies, we have identified 817 XBONG candidates with $L_X > 10^{42}$ erg s$^{-1}$ and X-ray to optical flux ratio $F_{XO} > 0.1$. Comparisons with WISE colors and NIR, optical, UV, and radio luminosities show that the loci of XBONGs are in-between those of control samples of normal galaxies and quasars and are consistent with low-luminosity quasars. We find that 43% of the XBONG sample have X-ray colors suggesting $N_H > 10^{22}$ cm$^{-2}$, double the fraction in the QSO sample, suggesting that a large fraction of XBONG are highly obscured AGNs. However, ~50% of the XBONGs are not obscured and have X-ray colors harder than those of normal galaxies. Some of these XBONGs have spatially extended X-ray emission. These characteristics suggest that they may be unidentified galaxy groups and clusters. Using the X-ray luminosity functions of QSOs and galaxies/groups/clusters, we estimate the approximate fraction of extended XBONGs to be < 20%. We also assess the approximate fraction of XBONGs whose AGN signatures are diluted by stellar light of host galaxies to be ~30%, based on their redshift and deviation from the extrapolation of the QSO $L_X$-$L_r$ relation.

# 1. INTRODUCTION

X-ray bright, optically normal galaxies (XBONG) were discovered with the Einstein X-ray telescope (Elvis et al. 1981). Their X-ray emission is as luminous as that of typical AGNs ($L_X > 10^{42}$ erg s$^{-1}$), but they show no optical AGN emission lines. XBONGs are recently attracting attention on both theoretical and observational grounds due to the increasing number of their detection with Chandra and XMM-Newton (Fiore et al. 2000; Comastri et al. 2002; Georgantopoulos & Georgakakis 2005; Kim et al. 2006; Civano et al. 2016). However, their nature remains unknown.

Three possibilities have been advanced: (1) XBONGs could be intrinsically luminous but heavily obscured AGNs, where neither broad nor narrow emission lines escape. This scenario is interesting because they may be part of a missing population of hard X-ray sources necessary to explain the observed X-ray background emission (Gilli, Comastri & Hasinger 2007; Ueda et al. 2014; Hickox & Alexander 2018). (2) They could be AGNs, where the nuclear emission lines have been diluted by the bright starlight of the host galaxy (Moran et al. 2002), i.e., type 2 AGNs with stellar light bright enough to outshine the AGN signature. These galaxies are often called optically dull AGNs (OD AGNs). Fitriana & Murayama (2022) identified 180 OD AGNs in the COSMOS survey and suggested that the host galaxy dilution may explain ~70% of their sample. (3) They could be groups of galaxies with a large amount of intragroup medium (IGM), compared to the typical interstellar medium (ISM) in a single galaxy. These groups may not be recognized in typical optical observations because they are either poor or fossil groups resulting from past galaxy mergers (e.g., Ponman et al. 1994) and dominated by a single elliptical galaxy (e.g., Jones et al. 2003). This type of system has also been called 'X-ray over-luminous elliptical galaxy' (OLEG; Vikhlinin et al.1999) and 'isolated OLEG' (IOLEG; Yoshioka et al. 2004).

The ChaMP (Chandra Multiwavelength Project) study identified 21 XBONG candidates in a sample of 136 Chandra extragalactic sources (Kim et al. 2006). Except for two, these XBONG candidates showed no sign of intrinsic absorption, suggesting that the absorption model is not universal (see similar reports from Georgantopoulos & Georgakakis 2005 and Hornschemeier et al. 2005). However, these results are all based on small samples.

Thanks to the Chandra Source Catalog (Evans et al. 2010), we can now revisit the question of the nature of the XBONG emission with a ~50 times bigger sample of XBONG candidates. By cross-matching the CSC2[1] with optical and IR catalogs from large all-sky survey projects, Kim et al. (2023 – Paper I) identified nearly 1000 XBONG candidates, which are classified as galaxies by the SDSS optical spectroscopy. The CSC2 provides additional information that can be used to further characterize this sample: (a) The X-ray spectral shape (measured by either the hardness ratios or the absorbing $N_H$ column) can help identify obscured AGNs and discriminate between XBONG emission and the average spectrum of normal galaxies. (b) The spatial extent can further point to extended emission regions that may be dominated by hot gas. (c) Temporal variability may point to AGN emission. Investigating the X-ray spectral, temporal, and spatial characteristics and multi-wavelength properties of the XBONG candidates, we will address the origin and nature of this unknown population.

This paper is organized as follows. In Section 2, we describe the selection criteria for XBONG candidates. In Section 3, we compare the Chandra X-ray properties of XBONGs with normal galaxies and QSOs. We explore the multi-wavelength (radio to UV) properties of XBONG in section 4. We discuss the nature of XBONGs in section 5. Then, we present our conclusions in section 6.

---

[1] https://cxc.cfa.harvard.edu/csc/

Throughout the paper, we adopt the following cosmological parameters: $H_o$= 69.6 (km/s)/Mpc, $\Omega_M$= 0.286, and $\Omega_\Lambda$= 0.714.

## 2. XBONG Sample Selection

By crossmatching the CSC2 and SDSS (DR16) spectroscopic (SpecObj) catalogs, we find 9350 unique matches (see Paper I). Of these counterparts, ~80% are classified by their SDSS optical spectra as QSOs and ~20% as galaxies; a small number of stellar counterparts are also found but will not be discussed here. The SDSS spectral class is determined by fitting the set of the galaxy, QSO, and star templates and chosen on the basis of the $\chi^2$ statistics (see more details in the SDSS website[2]). We also apply the flag to include only the SDSS sources with no redshift warning flag (i.e., zWarning=0), ensuring that classification and redshift are reliable.

Figure 1 shows the QSO and galaxy sub-samples in the $L_X$ - $F_{XO}$ plane, where $L_X$ is the X-ray luminosity, derived from the broad-band (0.5-7 keV) flux from CSC2 and using SDSS redshifts. We define the X-ray to optical flux ratio $F_{XO}$ as in Maccacaro et al. (1988), using for $F_X$ the broad-band (0.5-7 keV) flux from CSC2, and for the optical flux the r-band magnitude from the SDSS database:

$$\text{Log } (F_{XO}) = \log (F_X) + 5.31 + r / 2.5.$$

As described in Paper I, the galaxies (left panel of Figure 1) follow a tight near-linear distribution in the $L_X$-$F_{XO}$ plane. Using the Bayesian approach to linear regression of Kelly (2007), the best-fit relation of the galaxy distribution (the red line in Figure 1) is:

$$(L_X / 2.8 \times 10^{42} \text{ erg s}^{-1}) = (F_{XO} / 0.1)^{1.02}.$$

This linear relation holds for about five orders of magnitude for both $L_X$ ($10^{39}$ – $10^{44}$) and $F_{XO}$ ($10^{-4}$ - 10). The width of the red strip in Figure 1 is narrow, less than one order of magnitude full width orthogonally to the best-fit line, and does not change as a function of $L_X$. The distribution of QSOs in the $L_X$-$F_{XO}$ plane (right panel of Figure 1) is entirely different, clustered at the upper right corner with high $L_X$ and high $F_{XO}$. There is no $L_X$ - $F_{XO}$ correlation in the QSO sample.

Among the spectroscopically classified galaxies, we select the XBONG candidates by applying the following conditions:

$$L_X > 10^{42} \text{ erg s}^{-1} \text{ and } F_{XO} > 0.1 \quad \text{for XBONGs}$$

This $L_X$ cut eliminates X-ray bright normal (non-AGN) galaxies, either giant elliptical galaxies with a large amount of hot gas or late-type galaxies undergoing star formation bursts, which can have $L_X$ as high as $10^{42}$ erg s$^{-1}$ (see Nardini, Kim, & Pellegrini 2022 and references therein). $F_{XO}$ = 0.1 is the lower limit for QSOs from Paper I, which shows that 95% of the spectroscopically classified QSOs have $F_{XO}$ > 0.1.

---

[2] https://www.sdss.org/dr16/

We define 'normal' galaxies as those fulfilling the following conditions:

$L_X < 10^{42}$ erg s$^{-1}$ and $F_{XO} < 0.1$     for normal galaxies

Although these criteria are rather conservative, there are in our sample 250 galaxies with $L_X \sim 10^{42}$- $10^{43}$ erg s$^{-1}$ but $F_{XO} < 0.1$. These galaxies are optically brighter by a factor of a few or more than the selected XBONGs with similar $L_X$. In Section 5, we propose that some of these objects are AGNs with optical spectra diluted by stellar light from the host galaxy (see section 5).

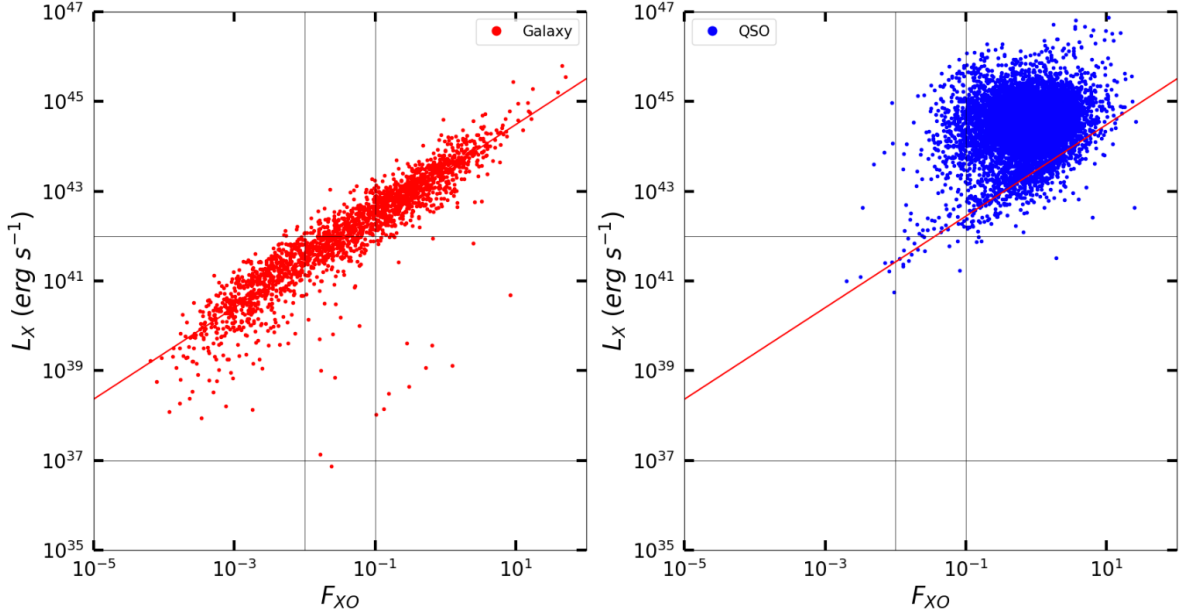

Fig. 1. $L_X$ against $F_{XO}$ (left) for galaxies and (right) QSOs from the SDSS spectroscopic sample. The red diagonal line is a near-linear relation (a slope of 1.02) which fits the galaxy sample. The vertical and horizontal lines identify the parameter spaces of XBONGs (galaxies with $L_X > 10^{42}$ erg s$^{-1}$ and $F_{XO} > 0.1$) and possible diluted AGNs ($L_X > 10^{42}$ erg s$^{-1}$, $F_{XO} < 0.1$) – see text for details. Also drawn are two additional lines (1) at $L_X = 10^{37}$ erg s$^{-1}$ below which stars dominate and (2) at $F_{XO}=0.01$ below which galaxies dominate with almost no QSO.

Given that we use the X-ray hardness ratios in our analysis, we exclude CSC2 sources for which the hardness ratios (HR=(X1-X2)/(X1+X2), where X1 and X2 are two of the CSC2 photometric energy bands; see section 3.2) cannot be determined. This excludes 7% of normal galaxies and 4% of XBONGs and QSOs. We retain sources with extreme hardness ratios (-1 or 1) where there is a significant detection, hence flux measurement, in only one of the photometric bands used in the hardness ratio. Our final samples include 865 normal galaxies, 817 XBONGs, and 6967 QSOs.

Some CSC sources have low significance and poor likelihood. To check how these low-quality sources affect our results, we have built high-significance X-ray samples by applying two additional selection criteria: significance > 3 and likelihood_class=TRUE, in the CSC2 master table[3]. This selection removes 43%, 15%, and 16% of the normal galaxies, XBONGs, and QSOs, respectively. We repeated all the tests in this paper using these reduced samples, finding no significant difference in the reported results. To quantitatively demonstrate this fact, we compare the results with/without poor-quality CSC sources in Table 3. This is because the objects in the original extended samples already have optical counterparts, which were classified by means of optical spectroscopy; hence they are likely real sources. This paper presents the results with no exclusion based on the X-ray source quality.

We primarily rely on the SDSS spectral class to separate galaxies, QSOs, and stars. The SDSS spectral catalog also provides a subclass based on spectral line diagnostics[4]. The subclass includes Star-forming (SF), Broadline, AGN, and N/A. The SDSS subclass AGN includes both LINERs and Seyferts. Because LINERs are mostly LLAGNs (e.g., Flohic et al. 2006; Gonzalez-Martin et a. 2009), we expect that most LINERs are included in the normal galaxy sample, but not in the XBONG sample ($L_X > 10^{42}$ erg s$^{-1}$). The subclass N/A indicates any object with no significant emission lines - most of them are likely early-type galaxies.

We show the $L_X$ - $F_{XO}$ relations for each sub-class in Figure 2 (the left 2x2 set is for class = Galaxy and the right 2x2 set is for class=QSO.) The $L_X$ – $F_{XO}$ relation is almost identical in different subclasses of a given class. All subclasses in class=Galaxy show a tight correlation between $L_X$ and $F_{XO}$, while all subclasses in class=QSO show no correlation and are scattered at high $L_X$ and $F_{XO}$. The only possible exception is subclass=Star-forming in class=QSO, which may show a correlation, but their $F_{XO}$ (0.1-10) is as high as the other QSOs.

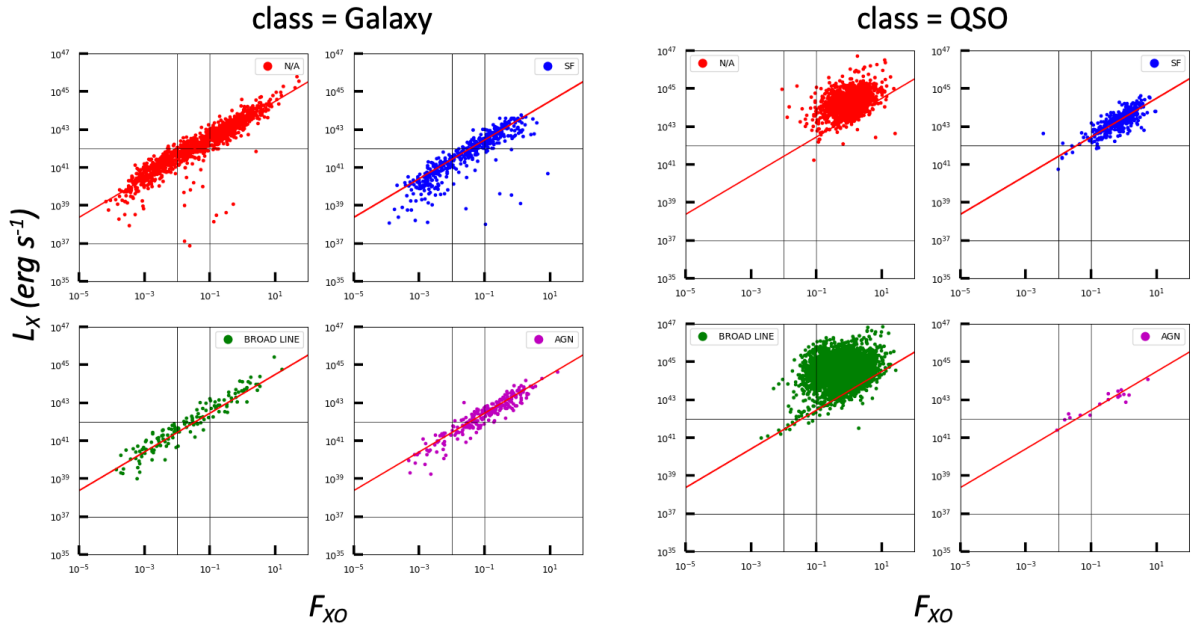

Fig. 2. Same as Fig 1, but each subclass is separately plotted (left) for class = Galaxy and (right) class = QSO. In each panel, we specify the subclass: N/A (not specified), SF (=star forming), BROAD LINE (broad emission lines), and AGN (narrow nuclear emission lines including LINERs). The red diagonal line is the best-fit relation of galaxies as in Figure 1.

---

[3] https://cxc.cfa.harvard.edu/csc/columns/significance.html
[4] https://www.sdss.org/dr16/spectro/catalogs/#Objectinformation

17% of the normal galaxies and 19% of the XBONGs have subclasses that indicate an AGN signature (subclass = AGN or Broadline). Among XBONGs (the upper-right quadrant in the left panel of Figure 2), the ratio of AGN to Broadline is 3 to 1. As seen in Figure 2, objects with the same class but different subclasses do not differ significantly in terms of the ensemble properties of $L_X$ and $F_{XO}$. More specifically, XBONGs with subclass=AGN or Broadline are not different from those with subclass =N/A or SF. Nonetheless, we repeated all the tests in this paper after excluding XBONGs with AGN signatures, finding no significant difference in the results. Here we present the results from the analysis of the total sample, regardless of the subclass. However, in some relevant cases (e.g., the statistical significance of spectral differences in Table 3, the measurement of the fraction of obscured XBONGs in Table 4, and the WISE color-color plots of each subclass in Figure 13), we report our measurements both for the total sample and for the sample that excludes AGN subclasses.

In Figure 3, we show the redshift distributions of three subsamples. The normal galaxies are found mostly at low z, with mean(z) = 0.1 and the 5% - 95% percentiles = 0.02 – 0.3. The XBONGs lie at higher z than normal galaxies, with mean(z) = 0.45 and the 5% - 95% percentiles = 0.1 – 0.9. The QSOs reach to highest z, with mean(z) = 1.5 and the 5% - 95% percentiles = 0.3 – 3.0. The z-distributions of XBONGs and QSOs are similar at low z (z < 0.5). At higher z, the number of XBONGs (per unit log(z) bin) significantly declines while the number of QSOs continues to increase until z ~ 2.

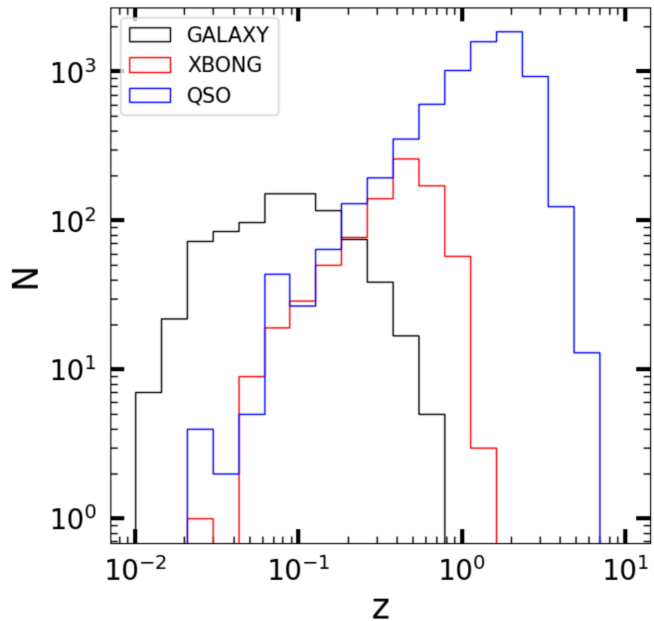

Fig. 3. The redshift distributions of the normal galaxy, XBONG, and QSO samples.

## 3. X-ray Properties of XBONGs

Compared with normal galaxies, XBONGs are simply an extension to higher $L_X$ and $F_{XO}$ (Figure 1 left). Compared with QSOs, XBONGs lie at the lower $L_X$ end in the $L_X$ – $F_{XO}$ parameter space occupied by QSOs (Figure 1 right). Tables 1 and 2 give the distribution of the samples in different

$L_X$ and $F_{XO}$ bins. By construction, the distributions of normal galaxies and XBONGs in this parameter space differ. The distributions of $L_X$ for XBONGs and QSOs also differ: most QSOs (80%) have $L_X > 10^{44}$ erg s$^{-1}$, while most XBONGs (92%) have $L_X < 10^{44}$ erg s$^{-1}$. In terms of $F_{XO}$, instead, QSOs, and XBONGs are similar, having $F_{XO}$ primarily between 0.1 and 10, although the fraction of QSOs with $F_{XO} > 1$ is slightly higher than that of XBONGs.

Table 1. Three sub-samples in different $L_X$ bins

|  | Normal galaxies | XBONGs | QSOs |
|---|---|---|---|
| Lx < 1e+40 | 103 (12%) | 0 | 1 |
| 1e+40 < Lx < 1e+41 | 285 (33%) | 0 | 1 |
| 1e+41 < Lx < 1e+42 | 477 (55%) | 0 | 22 |
| 1e+42 < Lx < 1e+43 | 0 | 336 (41%) | 173 ( 2%) |
| 1e+43 < Lx < 1e+44 | 0 | 415 (51%) | 1254 (18%) |
| 1e+44 < Lx | 0 | 66 ( 8%) | 5516 (79%) |
| total | 865 | 817 | 6967 |

Table 2. Three sub-samples in different Fxo bins

|  | Normal galaxies | XBONGs | QSOs |
|---|---|---|---|
| Fxo < 0.001 | 163 (19%) | 0 ( 0%) | 0 ( 0%) |
| 0.001 < Fxo < 0.01 | 452 (52%) | 0 ( 0%) | 11 ( 0%) |
| 0.01 < Fxo < 0.1 | 250 (29%) | 0 ( 0%) | 304 ( 4%) |
| 0.1 < Fxo < 1.0 | 0 ( 0%) | 615 (75%) | 4046 (58%) |
| 1.0 < Fxo < 10 | 0 ( 0%) | 193 (24%) | 2568 (37%) |
| 10 < Fxo | 0 ( 0%) | 9 ( 1%) | 38 ( 1%) |
| total | 865 | 817 | 6967 |

### 3.1 X-ray Hardness Ratios

We further compared the normal galaxy, XBONG and QSO samples, by making use of their X-ray spectral properties, as parameterized in the CSC2 hardness ratios. We used two hardness ratios from the CSC2[5] defined as

$$HR(ms) = \frac{F(medium) - F(soft)}{F(medium) + F(soft)} \quad \text{and} \quad HR(hm) = \frac{F(hard) - F(medium)}{F(hard) + F(medium)}$$

where F is the background-subtracted photon flux in each CSC2 energy band: soft (0.5-1.2 keV), medium (1.2-2 keV), and hard (2-7 keV). HR(ms), constructed with the fluxes in the medium and soft bands, takes advantage of Chandra's most sensitive energy range. HR(ms) is most effective to characterize softer sources like gas-dominated normal galaxies; it is also most sensitive to the amount of absorption in the range ($10^{20}$ cm$^{-2}$ < $N_H$ < $10^{22}$ cm$^{-2}$). HR(hm), constructed with the hard and medium bands, is sensitive to spectrally hard sources, like the power-law spectra of QSOs, and to larger amounts of absorption ($N_H$ > $10^{22}$ cm$^{-2}$). We use HR(ms) in Section 3.2 to compare the three sub-samples (normal galaxies, XBONGs, and QSOs) and use HR(hm) in Section 3.3 to compare XBONGs and QSOs and to investigate the amount of heavy obscuration.

---
[5] https://cxc.cfa.harvard.edu/csc/columns/spectral_properties.html

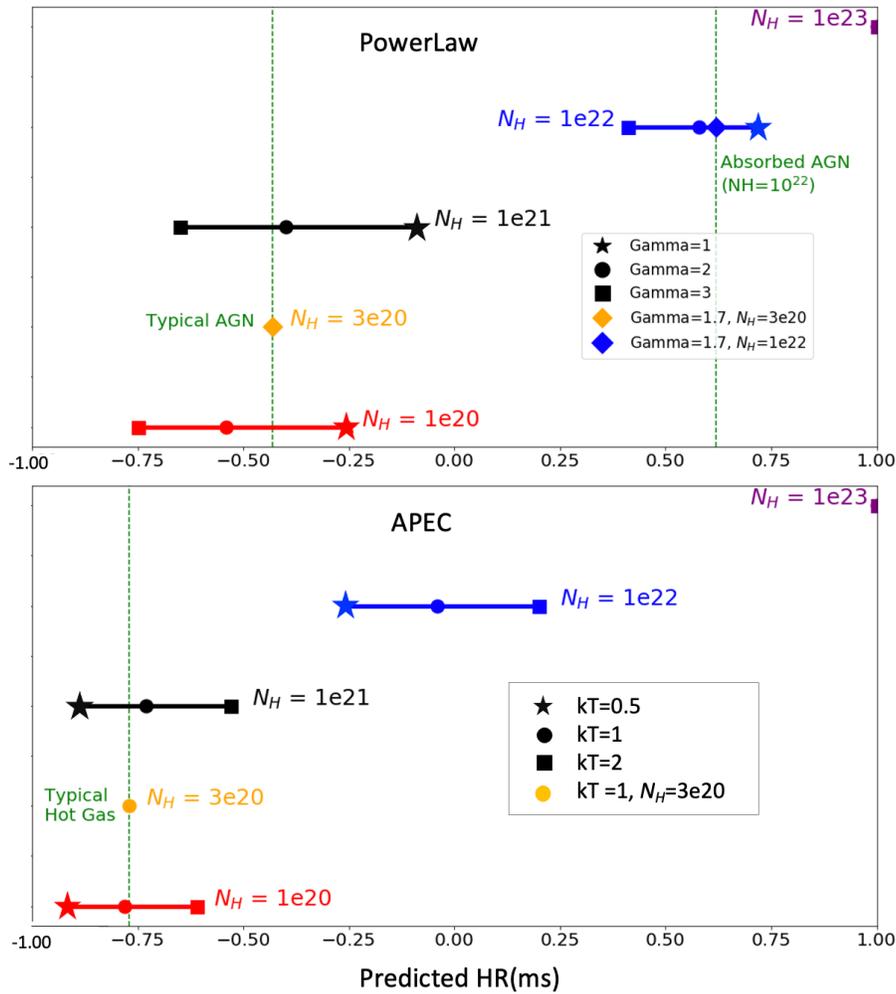

Fig. 4. Predicted HR(ms) for a range of emission/absorption model parameters. Marked by the vertical green dashed lines are the HR(ms) of the typical unabsorbed AGN (power-law slope of 1.7 and $N_H$=3 x $10^{20}$ cm$^{-2}$), absorbed AGN (power-law slope of 1.7 and $N_H$=$10^{22}$ cm$^{-2}$), and the emission of a hot-gas-rich galaxy (APEC with 1 keV gas and $N_H$=3 x $10^{20}$ cm$^{-2}$).

## 3.2 HR(ms) Distributions

Figure 4 shows the HR(ms) values calculated for a range of emission and absorption model parameters. In the top panel, a power-law model is used to represent the emission from QSOs, and in the bottom panel, an APEC model is used to represent the emission from the hot gas in galaxies. Different symbols (star, circle, and square) indicate different power-law slopes ($\Gamma$ = 1, 2, and 3) and different gas temperatures ($kT_{GAS}$ = 0.5, 1, and 2 keV). Different colors (red, black, blue, and purple) indicate different absorption column densities ($N_H$ from $10^{20}$ to $10^{23}$ cm$^{-2}$). The HR(ms) changes only slightly from $N_H=10^{20}$ cm$^{-2}$ to $N_H=10^{21}$ cm$^{-2}$, then significantly increases with increasing $N_H$ at $N_H > 10^{21}$ cm$^{-1}$. The HR(ms) of a typical unabsorbed AGN (power-law with $\Gamma=1.7$ and $N_H=3 \times 10^{20}$ cm$^{-2}$) and of an absorbed AGN ($\Gamma=1.7$ and $N_H=10^{22}$ cm$^{-2}$) are marked by the two green vertical lines in the top panel of Figure 4. The HR(ms) of a typical gas-rich galaxy (APEC with kT=1 keV and $N_H=3 \times 10^{20}$ cm$^{-2}$) is indicated by the green vertical line in the bottom panel. These three lines will be overlaid in all HR(ms) distribution figures to guide the expected locations of the typical AGNs, obscured AGNs, and hot gas-rich galaxies.

Figure 5 shows the CSC2 HR(ms) distribution of normal galaxies. We select the number of bins, ranging from 10 to 20, to best visualize the characteristic trends of each sample depending on the sample size. The vertical axis (number of objects in each bin) is normalized such that max = 100 for better visibility, and the error is scaled accordingly. The number next to the object type in the legend box indicates the scaling factor. We also calculate the mean HR(ms) and the error of the mean in each bin based on the error on HR(ms) from CSC2.

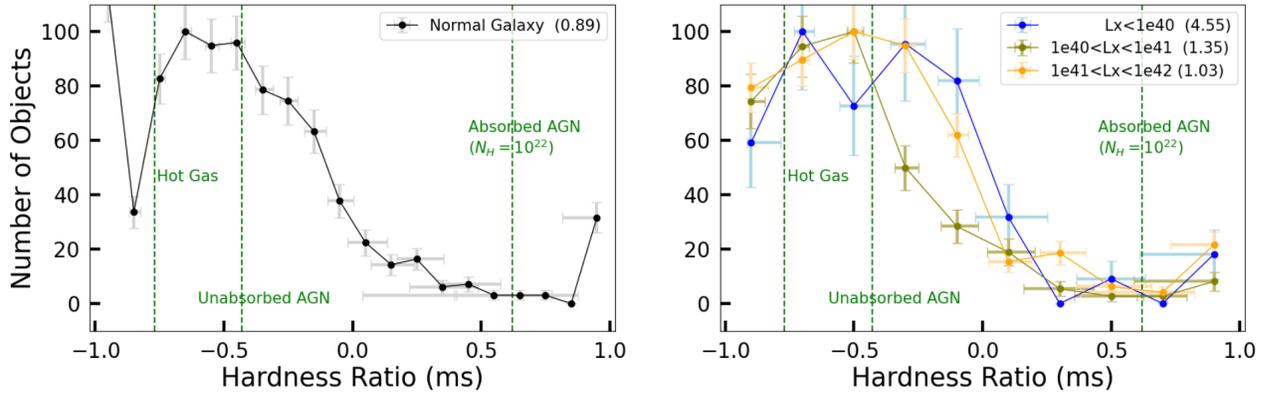

Fig. 5. (left) The HR(ms) distribution of the normal galaxy sample. (right) The normal galaxy sample is divided into three $L_X$ bins. The vertical green dashed lines correspond to those in Fig. 4. The number next to the object type in the legend box indicates the scaling factor of each distribution (see text).

As seen in the left panel of Figure 5 (the total sample), most normal galaxies have low values of HR(ms) (-0.7 < HR(ms) < -0.4), which can result from thermal models with kT=1-2 keV (Fig. 3). In galaxies dominated by gaseous emission, the temperature of the hot ISM and gaseous halos in E and S0 galaxies are typically kT<1.5 keV (Kim and Fabbiano 2015), and a similar temperature range is observed in starburst galaxies, although localized hotter regions may occur (see review in Fabbiano 2019). Energy injection from an AGN or past AGN activity can also increase the temperature of the ISM (Fabbiano & Elvis 2022). The deeper gravitational potential of galaxy groups is also reflected in higher values of kT (e.g., Fig 8 in Kim & Fabbiano 2015). In the hot-gas poor galaxies, the X-ray may be mainly from the populations of X-ray binaries (XRBs) with fairly hard X-ray emission (Fabbiano 1989, 2006; Boroson, et al. 2011).

In our CSC2-SDSS selected sample of normal galaxies, 13% have HR(ms) close to -1, which is consistent with hot gas emission with kT < 1.2 keV, leading to detection only in the CSC2 soft band. ~20% of the sample have HR(ms) between -0.3 and 0. Their spectra are too hard for typical emission from hot plasma in galaxies, so they could reflect either the XRB contribution or AGN or gravitational heating in addition to what would be expected in the typical dark matter halo of a normal galaxy. A larger value of the absorbing hydrogen column $N_H$ (as high as $10^{22}$ cm$^{-2}$) can increase the HR(ms), but this amount of obscuration, while observed in nuclear regions, is unlikely for the entire galaxy body. Higher values of $kT_{GAS}$ (as high as those in typical clusters) would be reflected in larger HR(ms) values; however, rich galaxy clusters are excluded by our selection criterion of $L_X < 10^{42}$ erg s$^{-1}$.

We find that there is an $L_X$ dependence on the HR(ms) values, as shown by the right panel of Figure 5, where we divide the normal galaxy sample into three $L_X$ bins, (1) $L_X < 10^{40}$ erg s$^{-1}$ (blue points), (2) $L_X = 10^{40} - 10^{41}$ erg s$^{-1}$ (olive points), and (3) $L_X > 10^{41}$ erg s$^{-1}$ (orange points). Galaxies in the low $L_X$ bin (blue points) have a spread of HR(ms) consistent with both a fraction of soft ~1 keV hot ISM dominated emission but also with harder spectra, as it would be expected by a substantial contribution of hard XRB emission in this luminosity range (e.g., Fabbiano 1989, 2006; Boroson, Kim & Fabbiano 2011). The fraction of normal galaxies with high HR(ms) (> -0.3) is smaller in the mid-$L_X$ bin ($10^{40} - 10^{41}$ erg s$^{-1}$; olive points), suggesting that a larger fraction of their average X-ray emission may be dominated by hot ISM and halos (see Boroson, Kim & Fabbiano 2011). The most luminous galaxies ($L_X = 10^{41} - 10^{42;}$ erg s$^{-1}$, orange points) in our sample have a distribution of HM(ms) similar to that of the lowest luminosity bin. Since XRB populations are unlikely to dominate their emission, these hard spectra may suggest either AGN contamination or associated larger dark matter halos.

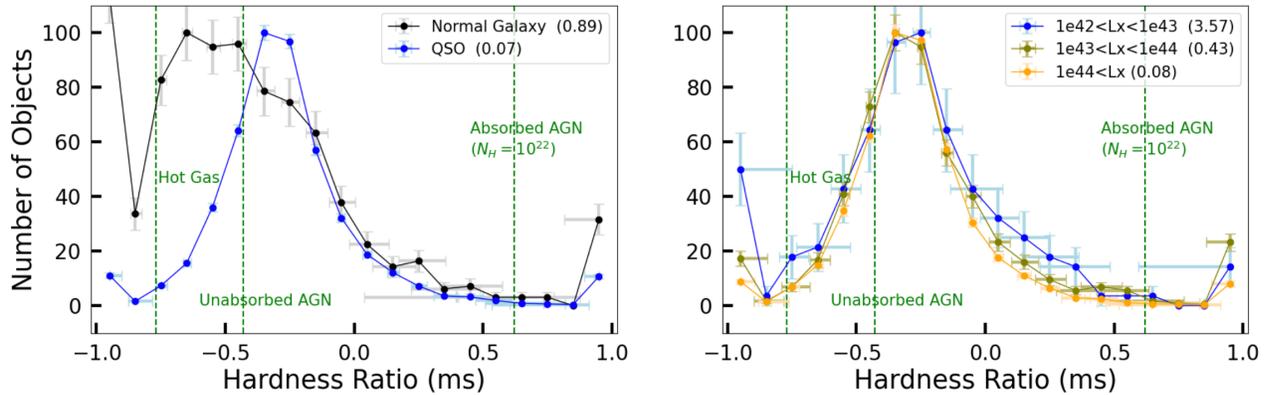

Fig. 6. (left) The HR(ms) distribution of the QSO sample in blue. The black points for the normal galaxy sample are the same as in Fig 5. (right) The QSO sample is divided into three $L_X$ bins. The vertical green dashed lines correspond to those in Fig. 4. The number next to the object type in the legend box indicates the scaling factor of each distribution.

We compare the HR(ms) distributions of the QSO and normal galaxy samples in Figure 6. The distribution of HR(ms) of QSOs is more skewed toward higher HR(ms) values than that of normal galaxies (the left panel of Fig. 5). The QSO distribution peaks at HR(ms) values in the range -0.5 ~ -0.2, which roughly correspond to those expected from an unabsorbed power law with $\Gamma = 1.7$ (Fig. 3). The HR(ms) distributions of the QSO and normal galaxy samples are remarkably similar at high HR(ms) > -0.2. In both cases, the fraction of objects at HR > 0 is small, suggesting that both samples primarily consist of unabsorbed populations. In the right panel of Figure 6, the HR distributions of QSOs in three $L_X$ bins are plotted – (1) $L_X = 10^{42} - 10^{43}$ erg s$^{-1}$, (2) $L_X = 10^{43} - 10^{44}$ erg s$^{-1}$, and (3) $L_X > 10^{44}$ erg s$^{-1}$. While there is no strong $L_X$-dependent trend, the QSO sample in the lowest $L_X$ bin ($10^{42} - 10^{43}$ erg s$^{-1}$; blue points) contains a relatively higher fraction of objects with higher HR(ms), between 0 and 0.5. This may indicate that an absorbed QSO population is most pronounced in the lowest $L_X$ bin.

We compare the HR(ms) distributions of the XBONG sample (red points) with the QSO and normal galaxy samples in Figure 7. The XBONG HR(ms) distribution is noticeably different from that of the normal galaxy sample, with overall larger values of HR(ms). It follows closely that of the QSO sample in the lower HR(ms) range, similarly peaking at HR = -0.3. However, the XBONG sample contains a higher fraction of objects with a high HR (>0) compared to the QSO sample, suggesting that the XBONG sample partially consists of an absorbed population. In the right panel of Figure 7 are plotted the HR(ms) distributions of XBONGs in three $L_X$ bins – (1) $L_X = 10^{42} - 10^{43}$ erg s$^{-1}$, (2) $L_X = 10^{43} - 10^{44}$ erg s$^{-1}$, and (3) $L_X > 10^{44}$ erg s$^{-1}$. There is no pronounced $L_X$-dependent trend.

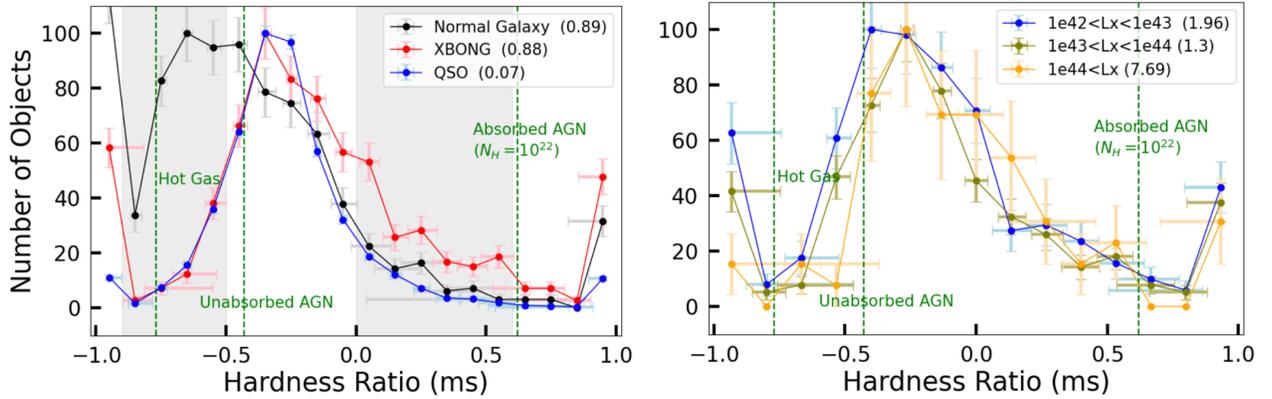

Fig. 7. (left) The HR(ms) distribution of the XBONG sample in red. The black (blue) points for the normal galaxy (QSO) sample are the same as in Fig 6. The shaded regions indicate the two HR(ms) bins where the statistical significance is calculated in Table 3. (right) The XBONG sample is divided into three $L_X$ bins. The vertical green dashed lines correspond to those in Fig. 4. The number next to the object type in the legend box indicates the scaling factor of each distribution.

To quantify the comparison of the three samples, we chose two HR(ms) bins (between -0.9 and -0.5 and between 0 and 0.6) where the significance of the difference between the three sub-samples is most pronounced. We calculated for each sample the fraction of objects in each bin, with their statistical significance. We then compared these fractions and calculated the significance of their difference. We did this comparison using both the full samples of galaxies and XBONGs and excluding those in the SDSS subclass of possible AGN signature (See Section 2). The results are summarized in Table 3. The HR(ms) bins, marked as shaded regions in the left panel of Figure 7, are the regions where the differences between the distributions are more evident. The low HR(ms) bin (-0.9 < HR(ms) < -0.5) contains 42% (± 3%) of the normal galaxy sample but only 10% (± 1%) of the XBONGs and 13% (± 1%) of the QSOs. The difference between the galaxy and XBONG fraction in this HR bin is highly significant, at the 10.4σ level. Similarly, the difference between the galaxy and QSO fraction is at the 9.6σ level. The difference between the XBONG and QSO fractions is instead at the 2.6σ level. Note that although XBONGs (Fig. 6, red points) and QSOs (Fig. 6, blue points) appear to follow each other for HR(ms) < -0.5, the XBONG fraction is lower than the QSO fraction because there are relatively more XBONGs at higher HR(ms).

Table 3. Significance of differences in the fractions of objects in two HR(ms) bins

| Samples compared | -0.9<HR(ms)<-0.5 | | | 0.0<HR(ms)<0.6 | | |
|---|---|---|---|---|---|---|
| | all (σ) | subclass (σ) | highQ (σ) | all (σ) | subclass (σ) | highQ (σ) |
| galaxy and XBONG | 10.4 | 9.2 | 8.6 | 6.6 | 5.5 | 6.6 |
| galaxy and QSO | 9.9 | 9.2 | 8.4 | 0.6 | 0.5 | 0.6 |
| XBONG and QSO | 2.6 | 1.7 | 1.8 | 7.1 | 6.0 | 7.2 |

column 'all': included all sources as defined in Section 2.
column 'subclass': excluded sources with possible AGN signatures in SDSS spectra (see Section 2).
column 'highQ': excluded poor-quality sources with significance < 3 or likelihood class=marginal (see Section 2)

The high HR(ms) bin (0 < HR(ms) < 0.6) contains 25% (± 2%) of the XBONGs, 9% (± 1%) of the normal galaxies, and 10% (± 0.4%) of the QSOs. The difference between the XBONG and normal galaxies fraction and that between XBONG and QSO fraction are both significant at the 6.6σ and 7.1σ level, respectively. The normal galaxies and QSO fractions are instead consistent (0.6σ significance of difference), suggesting that both samples do not comprise significant fractions of highly obscured objects.

In summary, XBONGs have HR(ms) significantly different from those of normal galaxies at the low HR(ms) values, where their HR(ms) distribution is marginally consistent with QSOs. XBONGs also have a significantly larger fraction of high HR(ms) objects than both normal galaxies and QSOs. The decreased fraction of XBONGs at low HR(ms) and the increased fraction of XBONGs at high HR(ms) may suggest that the XBONG sample includes a substantial number of obscured AGNs. From the fraction of XBONGs with HR(ms) > -0.2, the value beyond which the XBONG fraction exceeds that of the QSOs, we estimate that 50% of the XBONG sample may consist of obscured AGNs. HR(ms) = -0.2 corresponds to $N_H > 10^{21}$ cm$^{-2}$ for a 1.7 power-law spectrum, which exceeds the Galactic line of sight $N_H$.

## 3.3 HR(hm) Distributions

The HR(hm) (see the definition in Section 3.1) can be used to characterize hard power-law spectra and highly obscured sources. We show the distributions of HR(hm) for the XBONG and QSO samples in Figure 8. Because normal galaxies primarily emit X-rays below 2 keV, the hard (2-7 keV) band is not useful to characterize galaxies.

The trend observed in the HR(ms) distribution of Figure 7, that the fraction of obscured XBONGs is larger than that of obscured QSOs, is more clearly observed in the HR(hm) distribution. In the HR(ms) plot (Figure 7), the difference between the fraction of XBONGs and QSOs with high HRs is only significant in the range of HR(ms) = 0 – 0.6 (which corresponds to $N_H=10^{21}$ - $10^{22}$ cm$^{-2}$). In the HR(hm) plot of Figure 8, instead, the difference between XBONGs and QSOs is most pronounced for HR(hm) > 0.4, which corresponds to $N_H > 10^{22}$ cm$^{-2}$. 36% of the XBONGs have HR(hm) > 0.4 compared to only 12% of the QSOs. Combining both HR(ms) and HR(hm) distributions, we conclude that the fraction of obscured XBONGs is higher than that of QSOs with $N_H$ ranging at least to ~$10^{23}$ cm$^{-2}$.

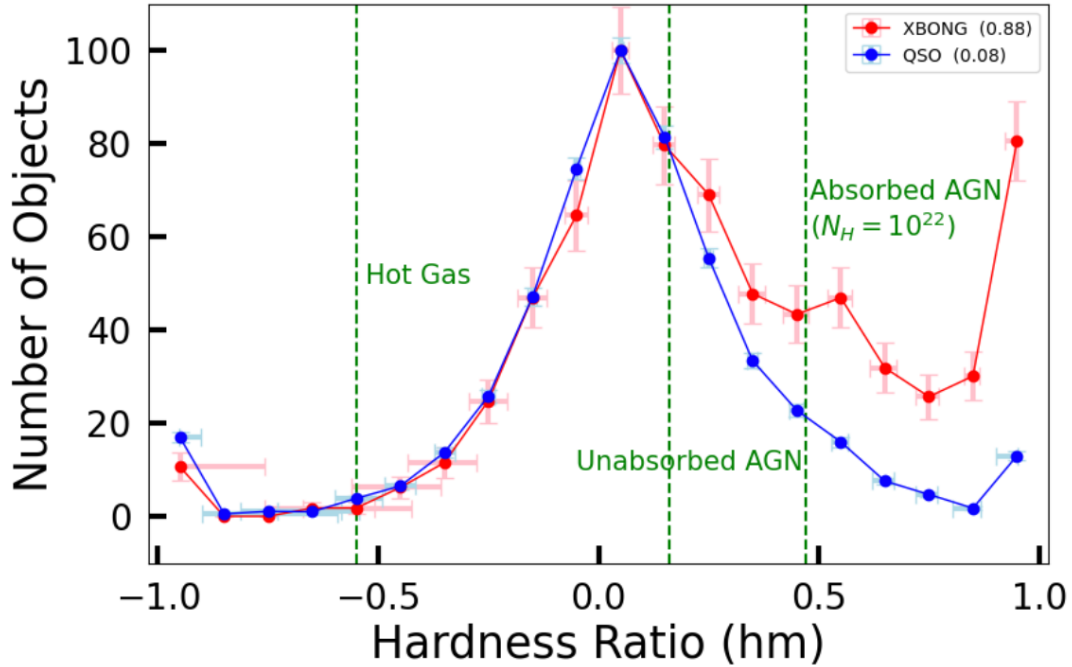

Fig. 8. Same as the left panel of Fig 7 for QSOs and XBONGs, but with the Hardness Ratio derived from the hard and medium CSC2 photometric bands.

The observed hardness ratios depend on the source redshift in such a way that a given amount of intrinsic obscuration would result in softer HR values for sources at higher redshifts. To examine redshift dependences, we have plotted the XBONG and QSO samples in the HR(hm) – z plane (Figure 9). The lines indicate power-law models with a slope of 1.7 and increasing $N_H$ from $10^{21}$ to $10^{24}$ cm$^{-2}$ (from bottom to top), with the thick line indicating the model with $N_H = 10^{22}$ cm$^{-2}$.

Figure 9 shows that a larger fraction of the XBONG sample (55%) lies above the dashed line of $N_H = 10^{21}$ cm$^{-2}$ (near HR(hm) = 0.2) than below this line, while the reverse is true for the QSOs, where 69% of the sample lies below this line. The discrepancy is stronger if we consider $N_H = 10^{22}$ cm$^{-2}$ (the thick line, a limit which is often regarded as obscured): for $z < 1.2$, where most XBONGs are found: 43% of XBONGs lie above this line, compared to 21% of QSOs. For $N_H > 10^{23}$ cm$^{-2}$, the obscured fractions are 12% and 3% for the XBONG and QSO samples, respectively.

Given the Chandra effective area[6], even the HR(hm) cannot sample enough data for extremely obscured objects ($N_H > 10^{24}$ cm$^{-2}$), and many sources above the $N_H = 10^{23}$ cm$^{-2}$ line in Figure 9 already have HR(hm) close to 1. We approximately estimate the fraction of possible sources with $N_H \gtrsim 10^{24}$ cm$^{-2}$ by applying HR(hm) > 0.99. This extreme obscuration is in the range of Compton-thick AGNs. This fraction is 9% for XBONGs and only 2% for QSOs. This can be considered a lower limit because more obscured AGN may not be detected with Chandra, or their luminosity is highly depressed so that we can see only the host galaxy and the nuclear reflection component that naively does not look obscured (Hickox & Alexander 2018; Fabbiano & Elvis 2022).

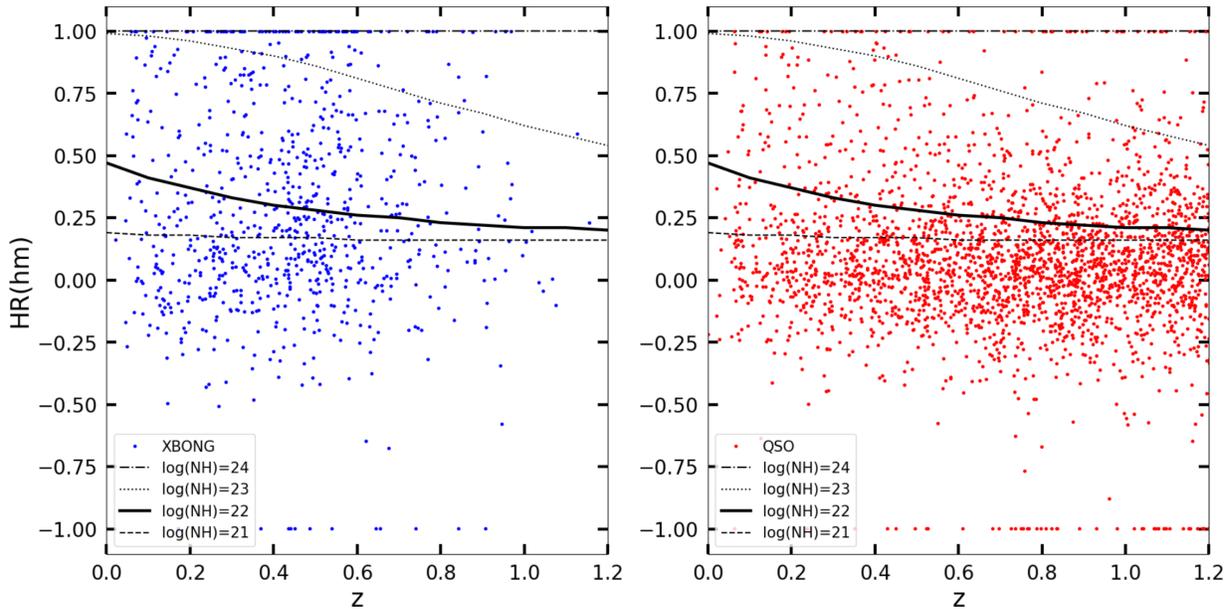

Fig. 9. HR(hm) is plotted against z for (left) XBONGs, and (right) QSOs. The lines indicate 1.7 power-law models $N_H = 10^{21}, 10^{22}, 10^{23}, 10^{24}$ cm$^{-2}$ from bottom to top.

We summarize the obscured fraction and the significance of difference in Table 4. The obscured fraction of XBONGs is roughly a factor of 2-4 higher than that of QSOs (at the ~7σ level). As stated in section 2, we only use the SDSS class (= GALAXY) to identify XBONGs, regardless of the SDSS subclass. Consistent fractions are calculated after excluding XBONGs with SDSS subclass=AGN or Broadline (listed in parentheses in Table 4).

---

[6] https://cxc.cfa.harvard.edu/proposer/POG/

We also looked at the HR(hs) distribution, made with the hard and soft bands. It also shows the increased fraction of XBONGs with higher HR values. Because of the limitation of the soft band, HR(hs) can trace only up to NH = $10^{22}$ cm$^{-2}$ as in HR(ms). We do not use HR(hs) in this paper.

Table 4. Obscuration fractions of XBONGs and QSOs

| | XBONG | QSO | significance of difference |
|---|---|---|---|
| $N_H > 10^{21}$ cm$^{-2}$ | 55.4% (52.7%)* | 31.9% | 6.6σ |
| $N_H > 10^{22}$ cm$^{-2}$ | 42.9% (40.8%)* | 21.2% | 7.4σ |
| $N_H > 10^{23}$ cm$^{-2}$ | 11.9% (10.8%)* | 2.7% | 6.9σ |
| $N_H > 10^{24}$ cm$^{-2}$ + | 9.3% ( 8.4%)* | 2.0% | 6.2σ |

\* The first number is calculated with all XBONGs with SDSS class=GALAXY, regardless of SDSS subclass. The number in the parenthesis is calculated after excluding XBONGs with SDSS subclass=AGN or Broadline.
+ They are approximately determined by HR(hm) > 0.99.

### 3.4 Source Extent and Variability

In addition to the X-ray spectral properties, we can use the spatial and temporal properties of the X-ray sources, which are also available in the CXC2. A CSC source is flagged as extended[7] if the PSF-deconvolved intrinsic size is determined with a significance of >5σ. About 9% of the entire sample is flagged as extended. Similarly, a CSC source is flagged as variable[8] if the flux varies within or between observations in one or more energy bands. To avoid unreliable cases caused by dithering across regions of uneven exposure (e.g., chip edges) during the observation, we exclude variable sources for which the dither warning flag is set. About 13% of the entire sample is flagged as variable.

Extended luminous X-ray sources are likely to be gas-dominated systems. X-ray variability would point to AGN emission. We explore this point in Figure 10, which shows the HR(ms) distributions of normal galaxies, XBONGs, and QSOs, using only '*extended*' sources (left) and '*variable*' sources (right).

In the extended normal galaxy sample, the fraction of objects with high HRs (between -0.5 and 0) is considerably reduced compared with the total sample. The distribution peaks near the green vertical line of the 1 keV hot gas, closely resembling the distribution of normal galaxies with a middle range of $L_X$ (see figure 5 for comparison with the full sample), which are dominated by hot gas with the least contamination by AGNs and X-ray binaries. The opposite trend is seen in the variable normal galaxy sample, where the peak of the distribution moves towards higher HR(ms) values, with an increasing fraction of higher HR(ms) objects, suggesting that the variable normal galaxy sample may have significant AGN contamination, or contain highly variable ultraluminous X-ray sources (ULXs; Kaaret, Feng & Roberts 2017)

---

[7] https://cxc.cfa.harvard.edu/csc/columns/srcextent.html
[8] https://cxc.cfa.harvard.edu/csc/columns/variability.html

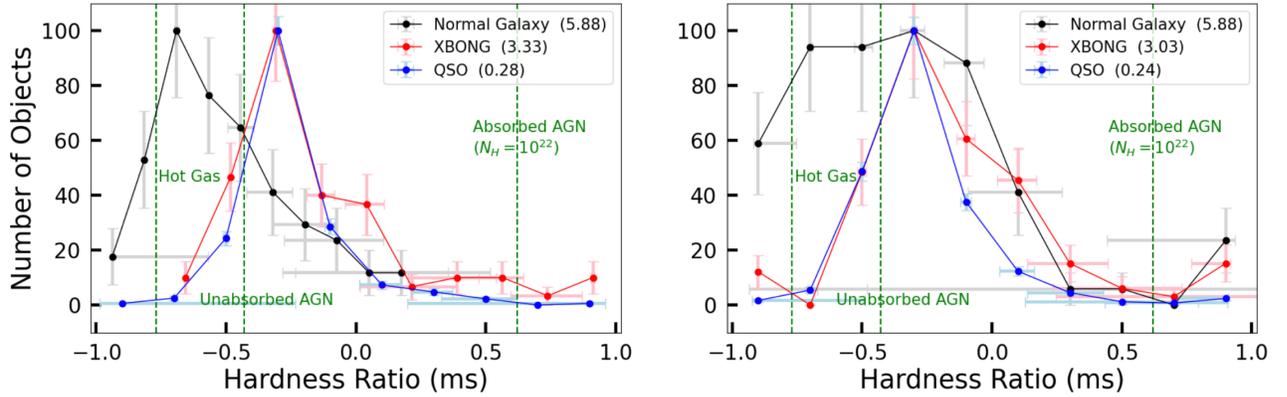

Fig. 10. Same as the left panel of Fig 7, but (left) with only sources flagged as extended in CSC2; (right) with only sources flagged as variable in CSC2.

The XBONG sample shows a similar trend as the galaxy sample but in a less dramatic way. Compared to the full XBONG sample (e.g., in Figure 7), the HR peak of the extended XBONG sample is sharper, and the HR distribution rapidly declines after the peak. Given the higher HR(ms) (and correspondingly higher kT), extended XBONGs are likely to be groups and clusters of galaxies.

We further examine the Chandra images of the XBONGs with the extended flag set. In Paper I, we have excluded the very extended sources, called convex hull[9], which include nearby galaxies and groups. For the remaining extended sources, we first select the objects with the raw-to-PSF size ratio > 5, then eye-examine their stack images. Although many are at large off-axis angles where the PSF is much larger, we find several X-ray groups and clusters among them. We show one example in Figure 11. This extended source, 2CXO J102155.7+344102, was detected at 14' from the aim point of a 5K observation. Compared with the corresponding Chandra PSF at the same location in the right panel of Figure 11, this X-ray source is undoubtedly extended. Its hardness ratio is HR(hm) = -0.1 (-0.2, 0.0), corresponding to no or little obscuration. Its WISE color (W1-W2) is 0.2, considerably lower than typical AGNs (>0.8) (see section 4). See more discussion in section 5.

In the variable sample (the right panel in Figure 10), the fraction of XBONGs with HR(ms) > 0 is higher than in the extended sample. They are likely to be obscured AGNs.

The QSO sample is least affected by the extended/variable flag (9%/13% of the QSO sample), likely because the nuclear X-ray emission dominates the QSO emission. The QSOs with extended_flag=TRUE may reflect the additional hot ISM/ICM of the host galaxy, possibly inside groups/clusters.

---

[9] https://cxc.cfa.harvard.edu/csc/dictionary/entries.html#convexhull

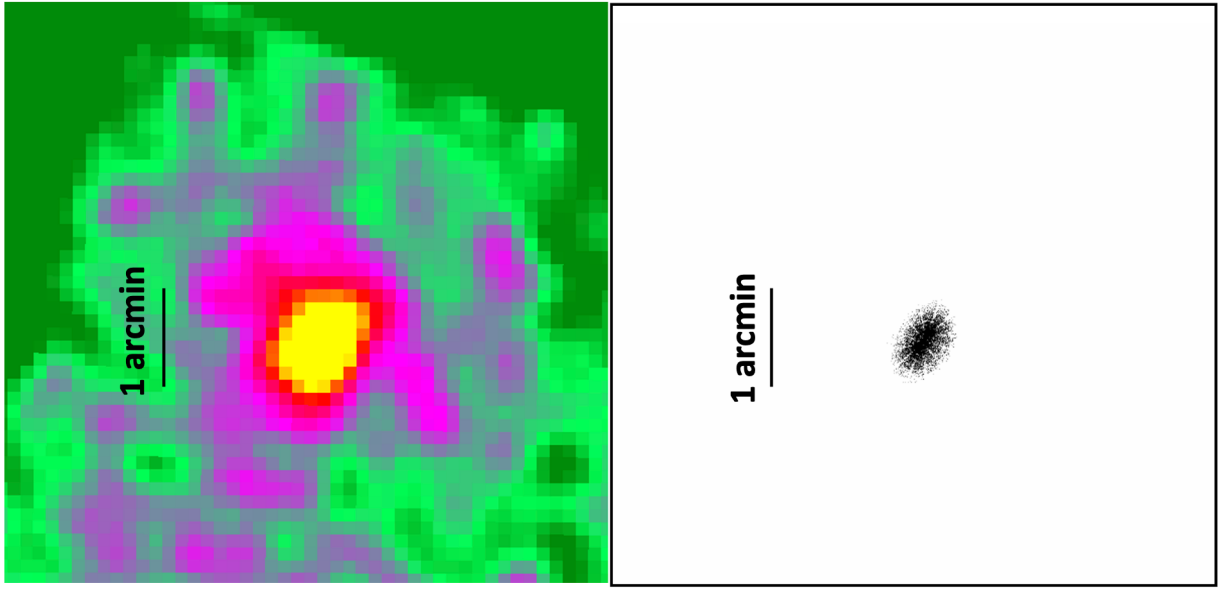

Fig. 11. An example of extended XBONG candidates. (left) 2CXO J102155.7+344102 at 14' from the aim point of a 5K observation. (right) The corresponding Chandra PSF at the same location. The vertical bar indicates 1'

The heavily obscured XBONGs can be best explored at higher energies (> 20 keV) than the Chandra energy band. We will pursue this study in the forthcoming paper with the hard X-ray catalog from the Swift/BAT (Oh et al. 2018) and INTEGRAL/IBIS (Bird et al. 2016 and Krivonos 2022) surveys.

We will also extend this study with the XMM-Newton source catalog (Webb et al. 2020), which provides more sources (0.2-12 keV), and the SWIFT XRT source catalog (Evans et al. 2020), which allows us to investigate variability (0.3-10 keV).

## 4. Multi-Wavelength Properties of XBONGs

We report here the results of multi-wavelength comparisons, including data ranging from the radio to the UV bands, used in addition to the X-ray data to further characterize the XBONGs and compare them with QSOs and normal galaxies.

### 4.1 IR Colors

As described in Paper I, the WISE color-color diagram is effective at separating galaxies and QSOs. In the left panel of Figure 12, we show the spectroscopically classified galaxies and QSOs in the $W_{12}$ - $W_{23}$ plane, where $W_{12}$ = W1 (3.4 μm) -W2 (4.6 μm) and $W_{23}$ = W2 (4.6 μm) -W3 (12 μm). They are extracted from the same samples shown in Figure 1 but restricted to objects detected with S/N > 2 in all three WISE bands. See Paper I (and Appendix A) for the details of crossmatching. There are 1325 galaxies and 4139 QSOs in Figure 12.

QSOs (plotted as blue points) are clustered at higher $W_{12}$ (0.6-1.7) and intermediate $W_{23}$ (2.3-3.8) colors. In contrast, galaxies (red points) are at lower $W_{12}$ colors (0-1) but span a wide range of $W_{23}$ (0-4.5). The blue horizontal line at $W_{12} = 0.8$ indicates the lower limit commonly used to identify QSOs (e.g., Jarett et al. 2011 and Stern et al. 2012). Above this limit, QSOs dominate with a small galaxy contamination (3%). To effectively identify galaxies with a small QSO contamination, in Paper I, we applied an upper limit of $W_{12} < 0.4$ for galaxies (the horizontal red line in Figure 12). Below this limit, galaxies dominate with a small QSO contamination (<10%). Also overplotted in Figure 12 is the AGN selection wedge (green dashed) from Mateos et al. (2012), which applies the $W_{12}$ limit as a function of $W_{23}$.

On the right panel of Figure 12, we separate the galaxies into normal galaxies and XBONGs (cyan points), as defined in Section 2. There are 638 normal galaxies and 460 XBONGs in this plot. Half of the XBONGs (52%) lie between the two horizontal lines, 14% are above the blue line ($W_{12} > 0.8$), and 34% below the red line ($W_{12} < 0.4$). Most XBONGs (79%) lie outside the AGN selection wedge of Mateos (2012). In terms of the WISE $W_{12}$ color, XBONGs are intermediate between normal galaxies and QSOs. In terms of the WISE $W_{23}$ color, XBONGs are similar to QSOs, except for a small fraction of XBONGs with low $W_{23}$ (<2) and an apparent lack of XBONGs with high $W_{23}$ (>3.5).

The $W_{12}$ color may be made bluer by absorption. This effect is more pronounced for less luminous AGNs (e.g., Stern et al. 2012). Interestingly, about half of the well-known Compton thick (CT) AGNs lie outside the AGN wedge or below $W_{12} = 0.8$ (e.g., Gandhi et al. 2015 and Boorman et al. 2019). These CT-AGNs are generally low in their X-ray (and IR) luminosity ($L_X < 10^{43}$ erg s$^{-1}$), while luminous AGNs dominate the mid-IR fluxes and are found at the expected location (inside the wedge or above $W_{12} = 0.8$). Similarly, type 2 AGNs are found more often below the AGN selection region than type 1 AGNs (Toba et al. 2014; Pena-Herazo et al. 2022). Regarding the mid-IR colors, the XBONGs are similar to the less luminous CT and Type 2 AGNs, consistent with our conclusion (Section 3) that a considerable fraction of the XBONG sample is obscured.

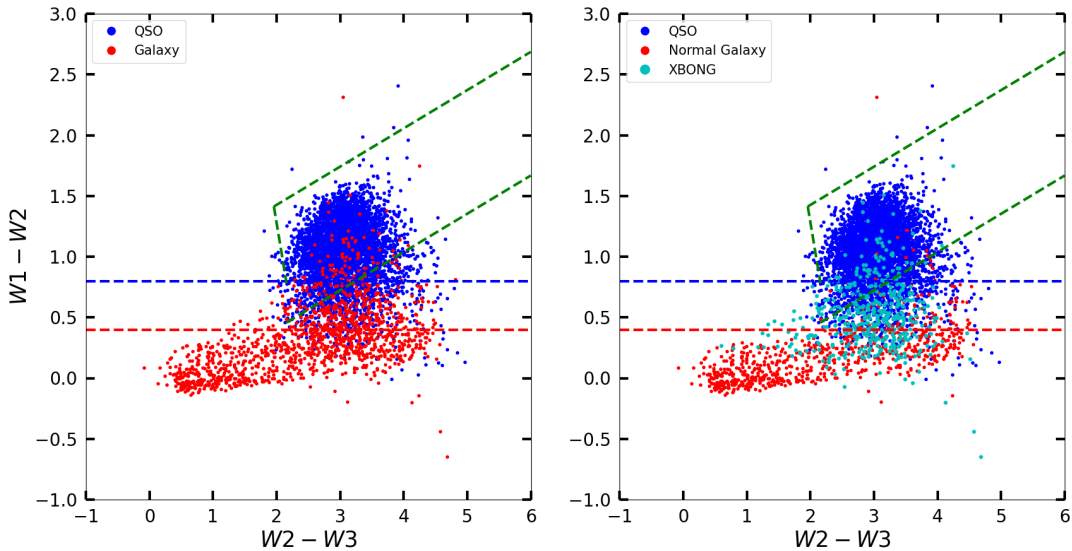

Fig. 12. (left) WISE color-color plot for the spectroscopically classified samples of galaxies (1325; red points) and QSOs (4135; blue points). (right) The galaxy sample is divided into two sub-samples, normal galaxies (638; red) and XBONGs (460; cyan). Two horizontal lines are overplotted to separate AGNs and galaxies at W1-W2 = 0.8 (blue dashed) applied by Stern et al. (2012) to select AGNs above this line and at W1-W2 = 0.4 (red dashed) applied by Kim et al. (2023) to select normal galaxies below this line. Also overplotted is the AGN selection wedge (green dashed) from Mateos et al. (2012). Many XBONGs lie outside the AGN wedge and between the two regions occupied by AGNs and galaxies.

However, a similar change in $W_{12}$ (becoming bluer) may be caused by the contamination of the stellar light from the host galaxy (e.g., Stern et al. 2012; Assef et al. 2013). The $W_{23}$ color distribution can provide additional information on the host galaxy effect. Regarding $W_{23}$, XBONGs are similar to QSOs but different from normal galaxies. The normal galaxies have a wide range of $W_{23}$ (0-4.5), depending on the star formation rate - early-type galaxies tend to have bluer $W_{23}$ than late-type galaxies. XBONGs lie in the middle, in a narrow range of W23 (2-3.5).

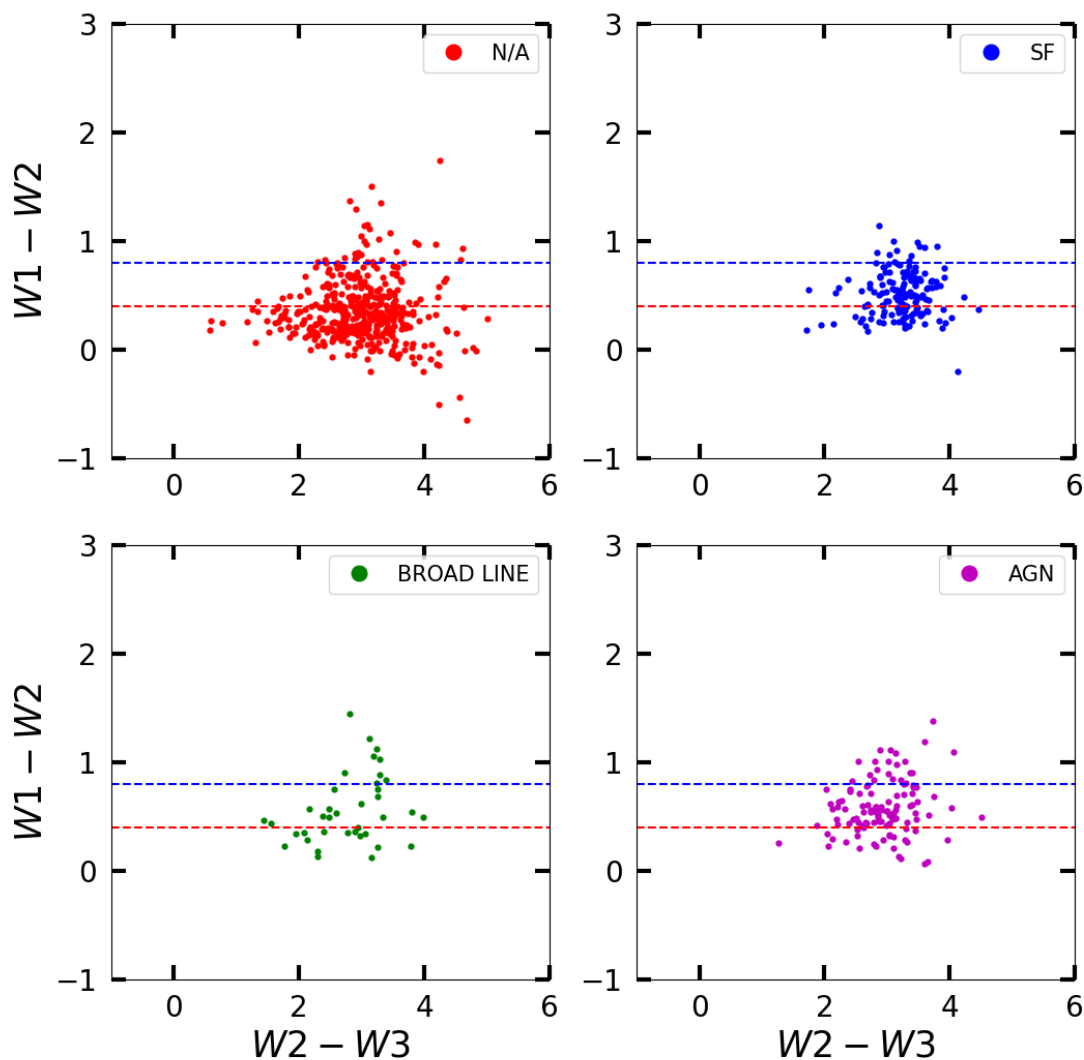

Fig. 13. Same as Figure 12, but showing only the XBONGs of each subclass.

Figure 13 shows each subclass of XBONGs in the WISE color-color plot to examine the difference between different subclasses. We find that a small fraction of XBONGs with AGN signature (subclass = AGN or Broad-line) exhibit no peculiar behavior. In W23, the AGN subclass is similar to the SF subclass, and the Broad-line subclass is similar to the N/A subclass. In W12, the AGN and SF subclasses are similar. The AGN subclass also includes LINERs (as described in section 2). Herpich et al (2016) reported that most LINERs fall below W12 < 0.8 because the host galaxy dominates the mid-IR emission (see also Mingo et al. 2016). The N/A subclass has a higher fraction below W12 < 0.4 than other subclasses. This is consistent with our results remaining unchanged with and without the inclusion of XBONGs with AGN signature.

We plot separately in Figure 14 the XBONGs classified as unobscured and obscured based on the X-ray hardness ratio. The two sub-samples are distributed similarly in the $W_{12} - W_{23}$ plane, both samples centered around ($W_{12}$ = 0.5 and $W_{23}$ = 3). However, two distinctions are identified. (1) The fraction in the AGN selection region with $W_{12}$ > 0.8 (or inside the wedge) is higher in the obscured than in the unobscured sample (20% vs. 8%). (2) The $W_{23}$ distribution of the unobscured sample is wider than the obscured (1.7-3.8 vs. 2.2-3.7 between 5 and 95% percentiles). This trend is more significant at $W_{12}$ < 0.4 (below the red line), where most galaxies are found.

In summary, the WISE colors indicate that XBONGs are in between QSOs and normal galaxies. The $W_{12}$ color suggests that some obscured XBONGs are similar to QSOs, but 80% are below the AGN zone, likely due to the obscuration. The unobscured XBONGs are also found below the AGN zone, but they may be also affected by the host galaxies, as seen in their wider $W_{23}$ distribution.

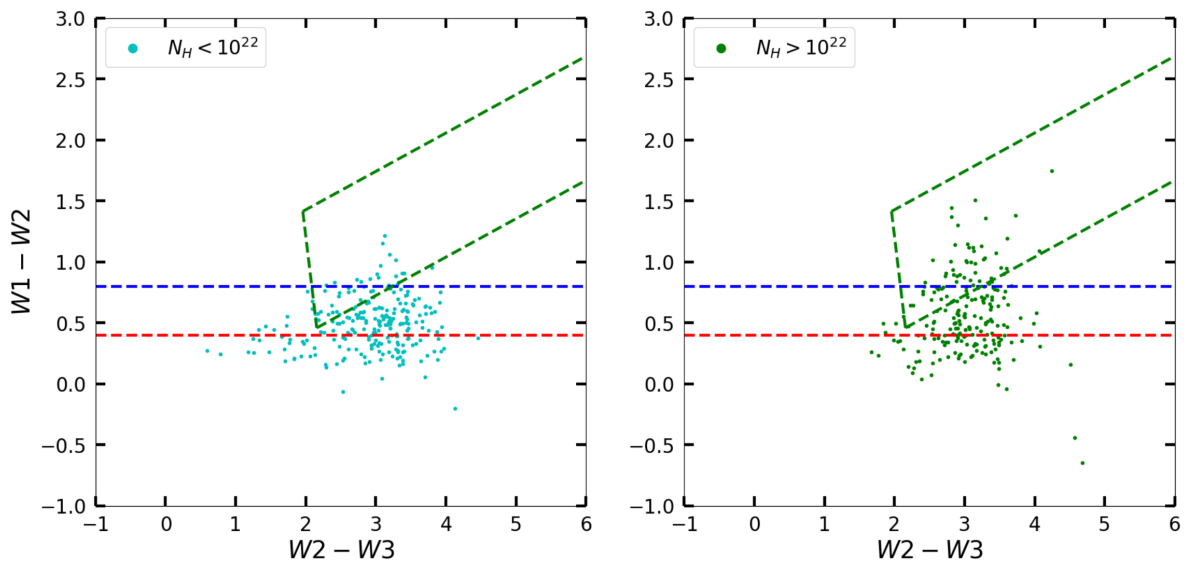

Fig. 14. Same as Figure 12 but only with (left) unobscured XBONGs ($N_H < 10^{22}$ cm$^{-2}$) and (right) obscured XBONGs ($N_H > 10^{22}$ cm$^{-2}$).

## 4.2 Optical and NIR Properties

We further explore the XBONG sample in the optical (r-band) and NIR (K-band) luminosities. In the top panel of Figure 15, we compare galaxies and QSOs. The parent samples are the same as in

Figure 1, but those in Figure 15 are also detected in each band under consideration. See Paper I (and Appendix A) for the details of crossmatching between different catalogs. The 2MASS K-band (2.2 μm) magnitude is converted to $L_K$ in the solar units, assuming that the absolute magnitude of the Sun is $M_{K\odot}$= 3.28 mag. Similarly, the SDSS r band magnitude is converted to $L_r$, taking the absolute solar magnitude $M_{r\odot}$= 4.68 mag.

The galaxy sample (red points in Figure 15) covers a narrow range of optical ($L_r$) and NIR ($L_K$) luminosity, approximately an order of magnitude. In contrast, the X-ray luminosity spans about three orders of magnitude. The correlations between $L_X$ and $L_r$ and between $L_X$ and $L_K$ are weak. In other words, $L_X$ does not depend on $L_K$ and $L_r$, both of which come from the stellar system (e.g., see Kim & Fabbiano 2013). This results in the near-linear relation between $L_X$ and $F_{XO}$, as seen in Figure 1 (see Paper I). The black dashed line shows the X-ray luminosity of low mass X-ray binaries (LMXB) for a given galaxy stellar mass represented by $L_K$ (Boroson, Kim & Fabbiano 2011) and indicates the lower limit of $L_X$ of galaxies.

On the contrary, in the QSO sample (blue points), there are strong correlations between $L_X$ and $L_r$ and between $L_X$ and $L_K$. This can be understood because the strong nuclei enhance their fluxes both in the optical/NIR and X-ray bands. In Paper I, we found the best-fit relation between $L_X$ and $L_r$ for QSOs: $(L_X / 1.6 \times 10^{43}) = (L_r / 10^{10})^{0.72 \pm 0.01}$. The best-fit relation between $L_X$ and $L_K$ is steeper, with a slope of $0.94 \pm 0.02$.

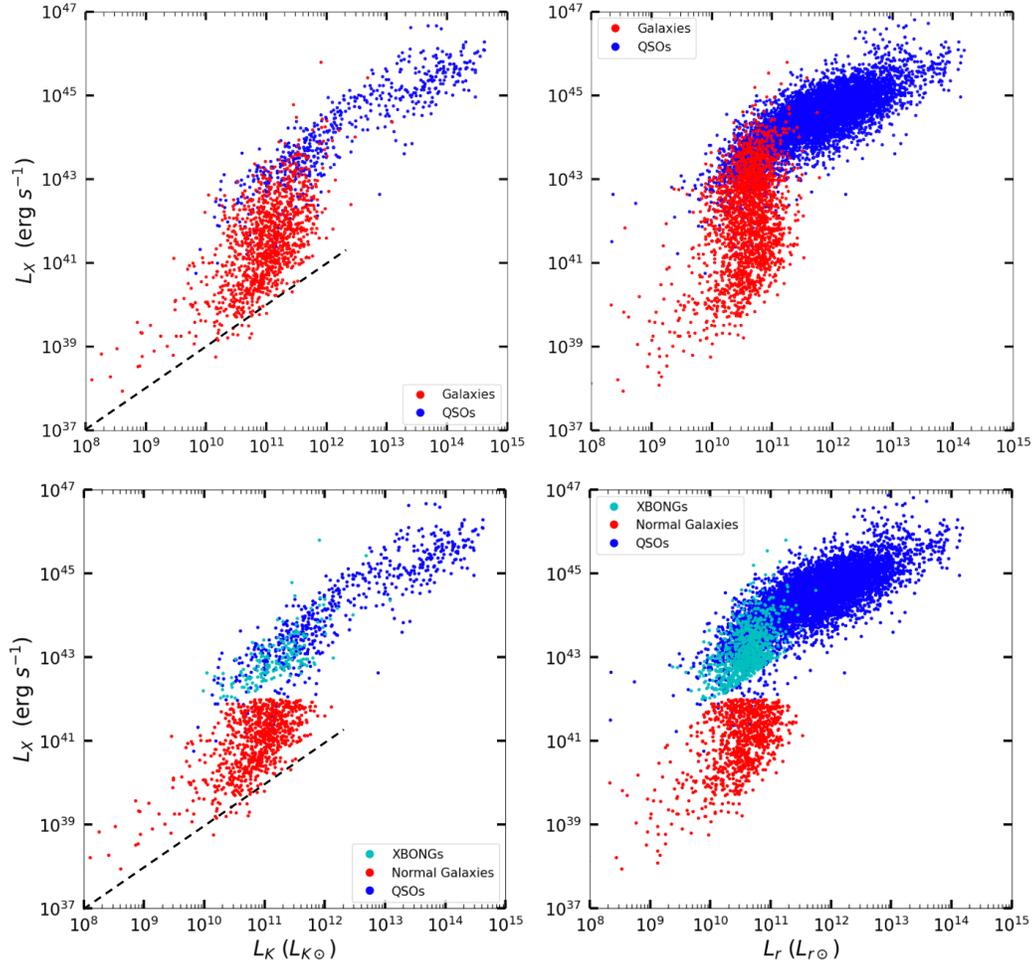

Fig. 15. (top )The X-ray luminosities of galaxies (red) and QSOs (blue) are plotted against (a) 2MASS K band, (b) SDSS r band luminosities. (bottom) The galaxies are separated into normal galaxies (red) and XBONGs (cyan). The black dashed line indicates $L_X$ from LXMBs for a given $L_K$ (Boroson, Kim & Fabbiano 2011).

In the bottom panel of Figure 15, galaxies are separated into normal and XBONGs, as described in section 2. Note that there is a gap between the two subsamples because of our XBONG selection criteria in section 2. The XBONGs (cyan points) lie between normal galaxies and QSOs in both plots of $L_X – L_r$ and $L_X – L_K$. In both plots, XBONGs appear to lie below the general trend of the QSO sample.

## 4.3 UV Properties

In Figure 16, we show a similar plot as in Figure 15, but with the GALEX[10] UV luminosities. The NUV (λeff = 2310A) and FUV (λeff = 1528A) data are taken from the revised GALEX UV catalog (Bianchi et al. 2017). To calculate the UV luminosities, we assume 8.53 mag for the absolute NUV of the Sun and 15.22 mag for the FUV.

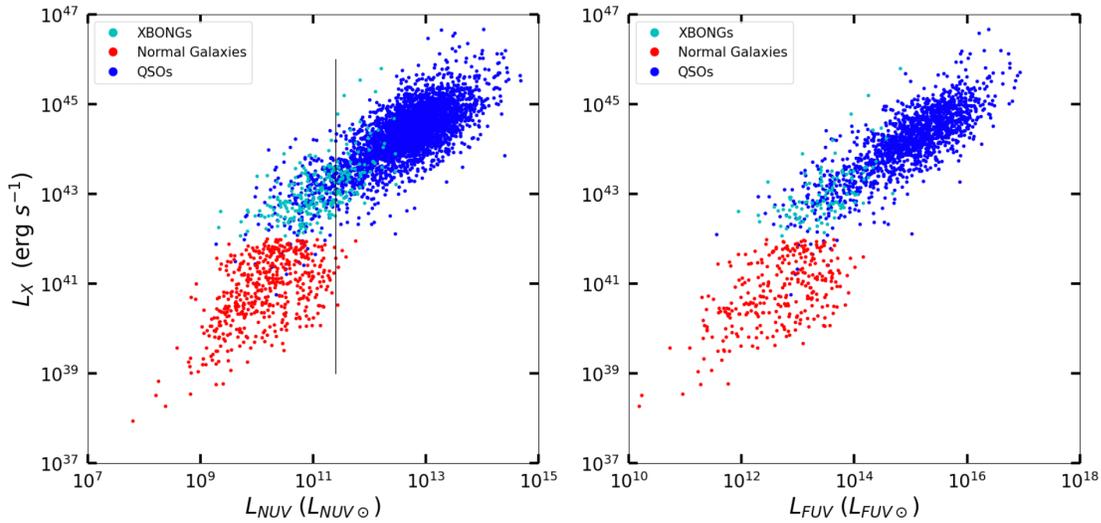

Fig. 16. Same as Figure 15, but with (a) GALEX NUV and (b) GALEX FUV luminosities. The vertical line indicates the maximum limit of normal galaxies, $L_{NUV} = 2.5 \times 10^{11} L_\odot$.

As for $L_K$ and $L_r$, there are poor correlations in the normal galaxy sample between $L_X$ and $L_{NUV}$ and between $L_X$ and $L_{FUV}$. In the QSO sample, there are strong correlations between $L_X$ and $L_{NUV}$ and between $L_X$ and $L_{FUV}$; these correlations are even tighter than those of the X-ray luminosity with $L_K$ and $L_r$. The strong X-ray - UV correlation for QSOs has been well-known since the early Einstein mission (Tananbaum et al. 1979). Its non-linear relation ($L_{2keV} \sim L_{2500Å}^{0.6}$) has been used in QSO cosmology (Risaliti & Lusso 2015).

---
[10] http://www.galex.caltech.edu/researcher/data.html

XBONGs (cyan points) are also overlaid in Figure 16. They lie between normal galaxies and QSOs in both plots at NUV and FUV and follow the same relation as the QSO sample. The best-fit relation of the QSO sample has a slope of 0.72 ± 0.01 (0.78 ± 0.01) for the $L_X$-$L_{NUV}$ ($L_X$-$L_{FUV}$) relation. The best-fit relation of the XBONG sample has a slope of 0.65 ± 0.05 (0.70 ± 0.09) for the same relation. The slopes in the two samples are consistent within 1.4σ (0.9σ).

The UV emission from AGNs is known to be related to the big blue bump (Elvis et al. 1994), and its detection critically depends on obscuration. The CSC2 hardness ratios show that most XBONGs detected in the UV are unobscured. Only 20% and 8% of 349 obscured XBONGs (with $N_H > 10^{22}$ cm$^{-2}$ in section 3.3) are detected in NUV and FUV, respectively. In contrast, 40% and 20% of 468 unobscured XBONGs (with $N_H < 10^{22}$ cm$^{-2}$ in section 3.3) are detected in NUV and FUV, respectively.

However, the presence of UV-bright ($L_{NUV} > 2.5 \times 10^{11} L_{NUV\odot}$ – more luminous than galaxies), unobscured ($N_H < 10^{22}$ cm$^{-2}$) XBONGs is puzzling. There are 56 such XBONGs in our sample. If they belong to the QSO family, why do they not exhibit typical AGN emission lines in their optical spectra? We further discuss these sources in Section 5.

## 4.4 Radio Properties

To explore the XBONG sample in the radio band, we use the combined radio catalog of Kimball & Ivezic (2008), which consists of FIRST (20cm), NVSS (20cm), WENSS (92cm), and GB6 (6cm). This catalog contains 2.7M entries in the region of the sky north of δ > -40°, which covers the areas of the SDSS survey. The radio emission can be either (1) AGN-driven radio cores and radio jets/lobes or (2) star formation-driven extended sources. QSOs will be dominated by the former and normal galaxies by the latter. We look for the radio characteristics of XBONGs in comparison with the other two samples.

In Table 5, we list the approximate numbers of counterparts for normal galaxies, XBONGs, and QSOs. Given the heterogeneous positional errors of the radio sources in the four radio surveys (the error is not provided in the combined catalog), we did not run the crossmatch tool. Instead, we searched for the radio counterparts within a few fixed separation radii. The number of matches increases rapidly from 1″ to 2″, but the increase rate is considerably flattened from 2″ to 3″, indicating that chance matches are preponderant for separation > 2″. We find that the chance to find a radio counterpart is highest for normal galaxies (19%) and lowest for QSOs (5%). The XBONG sample (12%) is intermediate. Because the different redshift ranges of the three subsamples may affect the result, we restricted the sample to QSOs within z<1.2, similar to the z range of the XBONG sample, in which case the QSO fraction is 6%, still significantly lower than that of XBONGs. Similarly, if we limit the XBONGs to have z<0.4, similar to the z range of the normal galaxy sample, the fraction is 13%, still significantly lower than that of normal galaxies. Considering that the fraction of radio-loud AGNs is ~10% (e.g., see a review by Panessa et al. 2019), the fraction of radio counterparts of the QSO sample is reasonable. The higher fraction of radio counterparts of the normal galaxy sample indicates that non-nuclear, SF-driven radio emission is present. XBONGs seem to be in the middle of the two comparison samples.

Using the matched list with separation < 2″, we plot $L_X$ against $L_{20cm}$ in Figure 17. We take the 20cm radio flux from FIRST, then from NVSS if no FIRST flux is available. The QSO sample

(blue points in Figure 17) has radio luminosity higher than the typical boundary of radio-loud and radio-quiet AGNs for a given $L_X$ (the thin black line in Figure 17, taken from Terashima and Wilson 2003). They show a tight correlation between $L_X$ and $L_{20cm}$ (blue line), $L_X \sim L_{20cm}^{0.7}$, consistent with the results from the analysis of the complete 3CR sample (Fabbiano et al. 1984). This slope is similar to that in the $L_X$ - $L_{UV}$ relation (section 4.3).

Table 5. The number of radio counterparts

|          | Normal<br>all | XBONG<br>all | XBONG<br>z<0.4 | QSO<br>all | QSO<br>z<1.2 |
|----------|---------------|--------------|----------------|------------|--------------|
| total    | 865           | 817          | 347            | 6967       | 2660         |
| sep < 1" | 107  12%      | 69   8%      | 35  10%        | 266  4%    | 125  5%      |
| sep < 2" | 165  19%      | 94  12%      | 46  13%        | 353  5%    | 164  6%      |
| sep < 3" | 179  21%      | 99  12%      | 46  13%        | 396  6%    | 178  7%      |

The normal galaxies (red points in Figure 17) lie an order of magnitude below the best-fit QSO correlation (blue line) extrapolated to normal galaxy luminosities. The SF-driven radio emission is known to correlate with $L_X$. Bauer et al. (2002) found a tight correlation ($L_X \sim L_{20cm}^{0.935}$) for the radio-emitting, star-forming galaxies in the Chandra deep field. Richards et al. (2007) also found $L_X \sim L_{20cm}^{0.95}$ for radio starbursts with $L_X < 10^{42}$ erg s$^{-1}$. We show this relation in Figure 17 (red line). The normal galaxy sample is consistent with this relation.

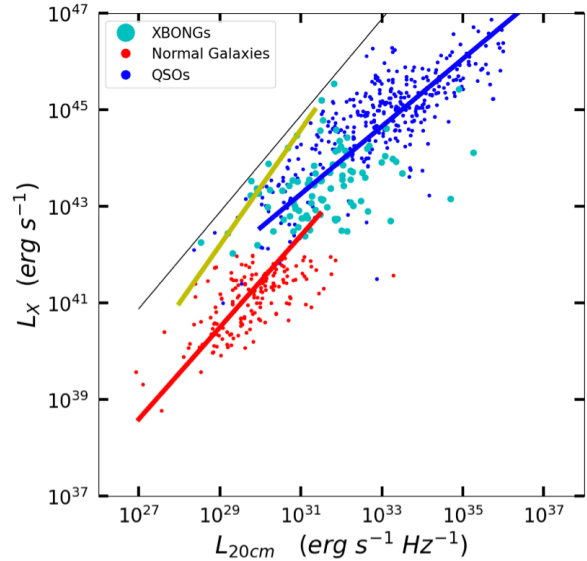

Fig. 17. The X-ray luminosities of normal galaxies (red), XBONGs (cyan), and QSOs (blue) are plotted against the 20cm radio luminosities. The blue line is the best-fit relation of QSOs, and the red line is from Richards et al. (2007) for SF galaxies with $L_X < 10^{42}$ erg s$^{-1}$. The thin black line indicates the boundary between radio-loud and radio-quiet QSOs (Terashima and Wilson 2003). The yellow line is the best-fit relation between $L_X$(ICM) of groups/clusters and $L_{20cm}$(AGN) of brightest cluster galaxies from Pasini et al. (2022).

The XBONGs are scattered between QSOs and normal galaxies. Most of the XBONGs could be consistent with the QSO correlation. One-third (two-thirds) are scattered above (below) the best-fit relation of QSOs (blue line). In contrast, about three-quarters (one-quarter) are scattered above (below) the extrapolated red line (star-forming galaxies).

The yellow line in Figure 17 indicates the best-fit relation between $L_X$ from the hot gas in groups/clusters and $L_{20cm}$ from AGNs of the brightest cluster galaxies (BCG), taken from Pasini et al. (2022). This relation is steeper ($L_X \sim L_{20cm}^{1-1.2}$) than the QSO relation and falls above the QSOs. A fraction of XBONGs with the highest $L_X$ for a given $L_{20cm}$ may be explained by groups/clusters (see section 5.2).

We also examined the 92cm WENSS and 6cm GB fluxes. They show similar trends as in Figure 17, albeit having a smaller (about 1/3) sample. Since the objects detected in either WENSS or GB are always detected in FIRST/NVSS, we do not show them separately.

In summary, we find that while a good fraction of the XBONG sample is consistent with the X-ray-radio relation of QSOs, there is a tail of XBONGs that may be consistent with the normal galaxies at low $L_X$ and with groups/clusters at high $L_X$.

## 5. The Nature of XBONGs

### 5.1. Obscured and Compton Thick AGNs

By definition, XBONGs are considerably more luminous in X-rays than normal galaxies but do not show any AGN line emission in their optical spectra. Both the X-ray spectral properties Sections 3.2 and 3.3) and the WISE colors (Section 4.1) suggest that at least half of the CSC2-SDSS XBONG sample consists of an obscured AGN population. Using the hardest CSC2 photometric band (Section 3.3), we estimated that the fraction of XBONGs with $N_H > 10^{22}$ cm$^{-2}$ is ~ 43%. This fraction is twice that of the CSC2 – SDSS QSO sample in the same redshift range (z < 1.2). The significance of the difference is 7σ. A similar but slightly higher fraction (52%) of the potentially obscured AGN population is obtained with the WISE W1-W2 color (see section 4.1).

Peca et al. (2022) analyzed the Chandra and XMM-Newton sources in the Stripe-82X field and measured the obscured AGN fraction, defined by $N_H = 10^{22}$ - $10^{24}$ cm$^{-2}$. They found that the obscured AGN fraction is 45-64% for X-ray sources with $L_X = 10^{42} – 10^{44}$ erg s$^{-1}$ and z <1, similar to the ranges of $L_X$ and z of our XBONG sample. Note that their sample includes both QSOs and galaxies. Given the $L_X$ range, their sample is comparable to our sample, with both XBONGs and QSOs included. Their obscured AGN fraction is similar to our measurement (43%) for the XBONG sample but somewhat higher than the 33% we derive if we consider the XBONG and QSO samples together. However, the CXC2-SDSS QSOs are mostly unobscured.

Given the limited energy range of the Chandra detector response, we can only measure the fraction of XBONGs with $N_H > 10^{23}$ cm$^{-2}$, and we take this fraction as a lower limit of Compton-thick AGNs ($N_H > 10^{24}$ cm$^{-2}$) because some of them cannot be detected with Chandra. When measured with HR(hm), this fraction of XBONGs is ~12%. This fraction is again significantly higher than that of the QSO sample (3%) in the same redshift range (z < 1.2). Considering all the objects with $L_X = 10^{42} – 10^{44}$ erg s$^{-1}$ and z = 0 - 1.2, the Compton-thick AGN fraction is about 5%: 4% from XBONGs and 1% from QSOs.

We note that the intrinsic X-ray luminosity of the Compton-thick AGNs is considerably higher than the observed $L_X$. The obscuration by $N_H=10^{24}$ cm$^{-2}$ can reduce $L_X$ by a factor of ~20 at z=0 or by a factor of ~4 at z=1. With the mean z of XBONGs being 0.4 ± 0.2, the most obscured AGNs in the XBONG sample can have intrinsic $L_X$ as high as $10^{45}$ - $10^{46}$ erg s$^{-1}$.

Given that the hypothesis that XBONGs are obscured AGNs can explain (at most) about half of this population, we will discuss other possibilities in the following sections.

## 5.2. Groups and Clusters of Galaxies

As described in Section 3, not all XBONGs can be explained as heavily obscured AGNs. About 15% of the XBONGs have X-ray spectra too soft for a typical AGN emission; 10% of these softer sources are CSC2 sources with significantly extended X-ray surface brightness. This is a lower limit on the number of intrinsically extended X-ray sources in the CSC2-SDSS XBONG sample. Other soft sources could also be intrinsically extended, although not detected as such by the CSC2 algorithms because they are faint and/or lie at large off-axis angles in the field of view, where the Chandra PSF is wider. Given the inferred temperatures of kT~1-2 keV and their X-ray luminosity, $L_X = 10^{42} – 10^{44}$ erg s$^{-1}$, these XBONGs could be associated with the emission of the large hot gaseous halos of groups and poor clusters of galaxies (see e.g., Figure 8 of Kim et al. 2015). Their optical counterpart may not be easily identified because optical catalogs are incomplete for poor groups/clusters, and the least massive or most evolved of these systems may only contain a handful of member galaxies. An extreme example is given by fossil groups and clusters dominated by a large X-ray luminous hot halo trapped in the gravitational potential of the merger that has evolved into a single galaxy (Ponman et al. 1994; Vikhlinin et al. 1999).

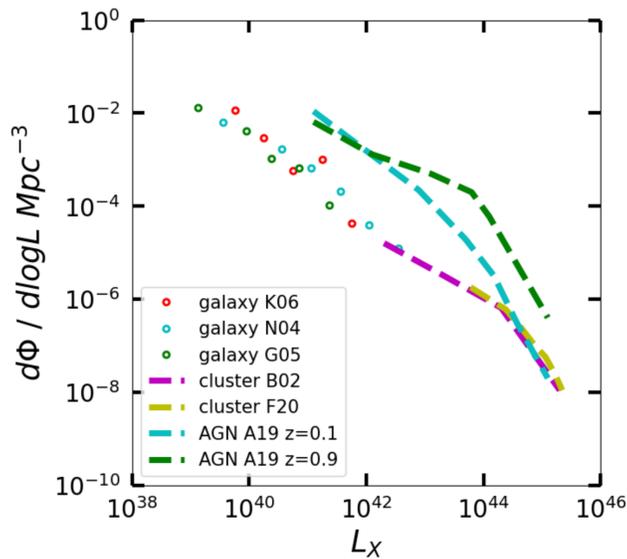

Fig. 18. X-ray luminosity functions of galaxies, clusters, and AGNs taken from the literature: K06 (Kim et al. 2006), N04 (Norman et al. 2004), G05 (Georgantopoulos et al. 2005), B02 (Bohlinger et al. 2002), F20 (Finoguenov et al. 2020), A19 (Annana et al. 2019).

We can obtain a rough estimate of the fraction of XBONGs that may be associated with groups and clusters of galaxies by comparing the X-ray luminosity functions (XLFs) of galaxies (e.g., Norman et al. 2004; Georgantopoulos et al. 2005; Kim et al. 2006), groups and clusters (Bohringer et al. 2002; Finoguenov et al. 2020), and AGNs at z=0.1 and z=0.9 (Ananna et al. 2019). These luminosity functions are shown in Figure 18. The different energy bands used by different authors were all converted to the Chandra broad band (0.5-7 keV). We did not plot the uncertainties on the luminosity functions for visibility, but in general, the errors are smaller than the differences in XLFs of different types. While no normal galaxies are expected in the interesting $L_X$ range (between $10^{42}$ erg s$^{-1}$ and $10^{44}$ erg s$^{-1}$), the expected fraction of groups/clusters to AGNs is approximately 10% in this luminosity range. Given that in the CSC2-SDSS sample, the number of XBONGs is about half that of QSOs in this $L_X$ range (Table 1), we conclude that extended groups and clusters can roughly account for only ~ 20% of the XBONG sample.

### 5.3. Dilution by Host Galaxies

Moran et al. (2002) suggested that some type 2 AGNs might be hidden if the stellar light of the host galaxy is bright enough to outshine the AGN signature and dilutes the nuclear emission lines. Some XBONGs may result from the same phenomenon, also suggested by the WISE colors in section 4.1. These objects are often called optically dull AGNs (OD AGNs). To try estimating the fraction of XBONGs that may be explained by dilution, we apply three conditions: (1) a redshift z>0.3 so that the aperture used in spectroscopic observations would include a large portion of a host galaxy; (2) an optically bright host galaxy, relative to the expected AGN optical luminosity for a given $L_X$; and (3) an unobscured AGN.

The redshifts of the XBONG sample range from 0.02 to 1.6 with mean z = 0.44 ± 0.2 (see Figure 3). 70% of XBONGs are at z > 0.3, where 1 arcsec corresponds to ~5 kpc. For them, the optical aperture would include at least half of the stellar light from the host galaxy, and it may be difficult to identify their optical AGN emission because of the dilution by the host galaxy. Assuming that the X-ray emission is primarily from an AGN (the $L_X = 10^{42}$ erg s$^{-1}$ cut for the XBONG sample is an extreme value for normal, non-AGN galaxies), we estimate the expected optical AGN luminosity by extrapolating the $L_X$-$L_r$ relation of QSOs (section 4.2). As seen in Figure 19, most XBONGs with $L_X < 3 \times 10^{43}$ erg s$^{-1}$ (red horizontal line) lie below or to the right of the QSO best-fit line (green dashed), i.e., have higher optical luminosity $L_r$ than expected from the AGN.

The dilution effect would be most significant for distant XBONGs with lower $L_X$. The fraction of XBONGs at z > 0.3 and with $L_X < 3 \times 10^{43}$ erg s$^{-1}$ is 47%. Subtracting those with $N_H > 10^{22}$ cm$^{-2}$ (as derived in sections 3.3 and 5.1), we estimate the fraction of diluted XBONGs to be 27%, roughly comparable to the XBONG fraction that cannot be explained with obscured AGN after excluding the estimate for cluster/group counterparts in the previous section.

Two hundred and fifty X-ray luminous galaxies with $L_X > 10^{42}$ erg s$^{-1}$ but $F_{XO} < 0.1$ were not included in the XBONG sample (see section 2). While they are still on the near-linear relation of galaxies (in Figure 1), they are characterized by higher optical luminosities for a given X-ray luminosity. Their location in the $L_X - L_r$ plane is marked by the yellow triangle in the right panel of Figure 19. Out of these 250 galaxies, 101 are found at $z > 0.3$. A large fraction of them may be explained by the dilution effect.

As discussed in Section 4.3, 56 XBONGs are unobscured ($N_H < 10^{22}$ cm$^{-2}$) and more luminous in NUV than galaxies ($L_{NUV} > 2.5 \times 10^{11}$ $L_{NUV\odot}$). Since the high NUV luminosity may indicate the presence of AGN emission, these XBONGs are unlikely to have group or cluster counterparts. Applying the same conditions described above, we find that 21 of these UV-bright XBONGs can be explained by dilution. For 6 more UV-bright XBONGs, the SDSS subclass indicates AGN spectral signatures. We cannot easily explain the nature of the remaining 29 UV-bright XBONGs, which consist of about 3% of our XBONG sample. Some of them could be mismatched. As determined by simulations in Paper I, the false match rate in crossmatching SDSS and CSC2 catalogs is ≤ 5%, i.e., up to 41 (out of 817 XBONGs) may have wrong optical counterparts.

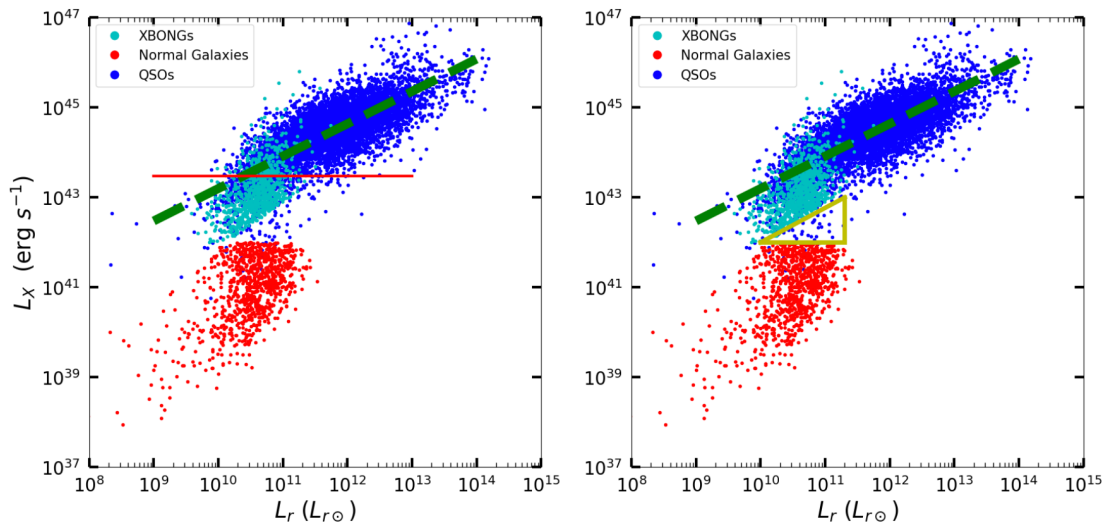

Fig. 19. (left) the same as the bottom right panel of Figure 15. The best fit $L_X$-$L_r$ relation of QSOs is overplotted (green dashed line). The red horizontal line indicates $L_X = 3 \times 10^{43}$ erg s$^{-1}$. Most XBONGs under this line are below the QSO best-fit line. (right) Also marked a gap between XBONGs and normal galaxies (see the text)

Finally, we consider the possibility that low-$L_X$ XBONGs may be explained by an inner radiatively inefficient accretion flow (RIAF) plus an outer radiatively efficient thin accretion disk (Yuan and Narayan 2004). While their optical and UV emission is suppressed, the inverse Compton emission can produce relatively strong X-rays ($L_X \sim 10^{42} - 10^{43}$ erg s$^{-1}$) from the hot RIAF. With a small sample of 48 OD-AGNs in the COSMOS field, Trump et al. (2009) suggested that the RIAF model might explain 30% (15 of 48) of the OD-AGNs with the highest $F_{XO}$, because the dilution by the host galaxy may not work in galaxies with high $F_{XO}$ (i.e., low optical fluxes for given X-ray fluxes). However, their RIAF candidates also have high $L_X$, 14 of 15 with $L_X > 10^{43}$ erg s$^{-1}$ and 6 with $L_X > 10^{44}$ erg s$^{-1}$, reaching the typical $L_X$ range of QSOs, too high for the RIAF model. This is because of the strong correlation among galaxies and XBONGs between $L_X$ and $F_{XO}$, as seen in Figure 1. Given the nature of RIAF, this model works at low accretion rates and hence at low $L_X$. Instead, some of their candidates may be obscured, assuming that unidentified groups/clusters are absent. We found that their RIAF candidates consist of a large fraction (66%) of their sample with high HRs (>0), in comparison to 22% with low HRs (<0). While the exact fraction of the RIAF candidates requires estimating the accretion rate with the reliable supermassive BH mass, we expect it to be limited to local (z<0.5), low-$L_X$ (< ~1 x $10^{43}$ erg s$^{-1}$) XBONGs, if not obscured and not diluted. As seen in Figure 19, there are very few XBONGs with low $L_X$ and high $F_{XO}$, i.e., below the red line and above the green lines. We note that the unexplained 3% of XBONGs (in the above paragraph) cannot be explained by the RIAFs because of their high NUV luminosity.

## 6. Conclusions

By cross-matching CSC2 X-ray sources and SDSS spectroscopic samples, we have identified 817 XBONG candidates and two control samples of (865) normal galaxies and (6967) QSOs. XBONGs are spectroscopically classified as galaxies but have very high X-ray luminosity ($L_X$ > $10^{42}$ erg s$^{-1}$) and high X-ray to optical flux ratios ($F_{XO}$ > 0.1), both of which are characteristics of QSOs. We reach the following conclusions by examining their X-ray spectral, spatial, and temporal properties and multi-wavelength properties.

- The X-ray hardness ratios indicate that a large fraction of the XBONG sample consists of a population of obscured objects. The fraction of XBONGs with a high CSC2 medium-to-soft band hardness ratio HR(ms), implying $N_H$ > $10^{21}$ cm$^{-2}$, is significantly higher than those of normal galaxies (at the 6.6σ level) and QSOs (7.1σ).
- Considering the redshift-dependent CSC2 hard-to-medium band hardness ratio HR(hm), the fraction of obscured XBONGs with $N_H$ > $10^{21}$ cm$^{-2}$ is 55%, and that with $N_H$ > $10^{22}$ cm$^{-2}$ is 43%. These obscured fractions are a factor of two higher than in the QSO sample. Roughly half of the XBONGs may be obscured AGNs.
- The obscured fraction of XBONGs with extremely high $N_H$ (> $10^{23}$ cm$^{-2}$) is 12%. Most of them are possibly in the range of Compton-thick AGNs with $N_H$ > $10^{24}$ cm$^{-2}$. The similar fraction of QSOs is very small (2%). Given the Chandra response function, these estimates are lower limits.
- In the WISE color-color plot, the normal galaxy (W1-W2 < 0.4) and QSO (W1-W2 > 0.8) lie in different locations. 52% of the XBONGs lie between AGNs and galaxies (0.4 < W1-W2 < 0.8). Based on their WISE colors, most (80%) of the XBONGs classified as obscured according to their X-ray hardness ratios lie outside the AGN selection region. The WISE colors of the X-ray unobscured XBONGs may be affected by those of the host galaxies.
- Comparing $L_X$ with luminosities in other bands (2MASS K, GALEX UV, and radio 20cm), we find that the XBONGs lie between normal galaxies and QSOs.
- XBONGs follow normal galaxies in the $L_X$-$L_r$ and $L_X$-$L_K$ relations and are intermediate between normal galaxies and QSOs in the $L_X$-$L_{20cm}$ relation.
- We find that 56 XBONGs follow a relation consistent with that of QSOs in $L_X$-$L_{UV}$. These XBONGs are UV-bright, indicating the lack of absorption, so it is puzzling that they do not show optical AGN emission lines. Only half of them could be explained with galaxy dilution of the AGN spectrum.

- The X-ray extent of some of the XBONGs with softer X-ray hardness ratios suggests that the XBONG sample also consists in part of a population of extended, X-ray soft objects. These XBONGs could be groups or poor clusters of galaxies, including fossil groups. Based on the previously determined X-ray luminosity functions of different classes of sources, we roughly estimate that less than 20% of XBONGs may be X-ray extended groups or clusters.
- Dilution by the stellar light of the host galaxies may explain the remaining XBONGs (~30%) if the $L_X$-$L_r$ relation of QSOs applies to XBONGs.

## ACKNOWLEDGEMENTS

We have extracted archival data from the Chandra Data Archive and the Chandra Source Catalog version 2. The data analysis was supported by the CXC CIAO software and CALDB. We have used the NASA NED and ADS facilities. This work was supported by the Chandra GO grants (AR1-22012X) and NASA contract NAS8–03060 (CXC).

Appendix A. CROSSMATCH

The crossmatching procedure is described in detail in Paper I. Here we briefly summarize the key steps for reader's convenience. To crossmatch the CSC2, we used the NWAY package v4.5.2[11] (Salvato et al. 2018). The NWAY parameters were set to optimize the matching fraction while minimizing false positives and negatives. (a) We accept only unique matches with separation < 3 arcsec, regardless of the positional error. (b) We set the minimum probability ratio for the secondary match *acceptable-prob* = 0.25 - smaller than the default value of 0.5 - to determine a 'unique match' conservatively. (c) We set *match_flag*=1 (indicating the most probable match). (d) We set the probability threshold (i.e., the probability that one of the associations is correct) *p_any* = 0.5. To evaluate the rate of false matches, we have run extensive simulations, following the same matching procedure outlined above but after shifting the source positions in eight directions (horizontal, vertical, and diagonal) by 30 arcseconds. We found that the rate of false matches is a strong function of source density. With our strict selection criteria, we determined the false match rate of ~5%.

---

[11] https://github.com/JohannesBuchner/nway